%% file: main_arxiv.tex
\documentclass{article}

\usepackage{arxiv}
\usepackage[utf8]{inputenc} % allow utf-8 input
\usepackage[T1]{fontenc}    % use 8-bit T1 fonts
\usepackage{amssymb}
\usepackage{float}
\usepackage[caption=false]{subfig}
\usepackage[export]{adjustbox}
\usepackage{graphicx}% Include figure files
\DeclareGraphicsExtensions{.jpg,.png,.pdf}%prefer jpg over png
\usepackage{bm}% bold math
\usepackage{nicematrix}
\usepackage{hhline}
\usepackage{tikz}
\usepackage[colorlinks=true]{hyperref}% add hypertext capabilities
\usepackage{cleveref}
\usepackage[all]{hypcap}% Hyperlink jumps to figure instead of caption
\usepackage{derivative}
\usepackage{xr}

\input{preamble}
\newcommand{\classiffontsize}{\tiny}%
\newlength{\tableJoinSep}
\graphicspath{{./jpgimages/}{./images/}}
\begin{document}
\input{arxiv/frontmatter}

\input{sections/introduction}

\input{sections/meaningfulness}
\input{sections/results}
\input{sections/conclusion}

%\data{The data that support the findings of this study are available at \url{https://github.com/lyc4038/MotifsInSelfOrganizingCells}}

\section{Supplementary Information}
\input{arxiv/SI}

\bibliography{main}

\end{document}

%% file: preamble.tex
\newcommand{\featAfull}{\vec{A}(n,t|\alpha_{\text{CIL}},\alpha_{\text{CF}})}
\newcommand{\featBfull}{\vec{B}(t|\alpha_{\text{CIL}},\alpha_{\text{CF}})}
\newcommand{\featCfull}{\vec{C}(\alpha_{\text{CIL}},\alpha_{\text{CF}})}
\newcommand{\featDfull}{\vec{D}(n|\alpha_{\text{CIL}},\alpha_{\text{CF}})}
\newcommand{\featEfull}{\vec{E}(n,t|\alpha_{\text{CIL}},\alpha_{\text{CF}})}
\newcommand{\featFfull}{\vec{F}(t|\alpha_{\text{CIL}},\alpha_{\text{CF}})}
\newcommand{\featGfull}{\vec{G}(n,t|\alpha_{\text{CIL}},\alpha_{\text{CF}})}
\newcommand{\featHfull}{\vec{H}(t|\alpha_{\text{CIL}},\alpha_{\text{CF}})}

\newcommand{\featAtbl}{\vec{A}(n,t|\vec{\alpha})}
\newcommand{\featBtbl}{\vec{B}(t|\vec{\alpha})}
\newcommand{\featCtbl}{\vec{C}(\vec{\alpha})}
\newcommand{\featDtbl}{\vec{D}(n|\vec{\alpha})}

%% file: arxiv/frontmatter.tex
\title{Motifs in Self-Organising Cells}% Force line breaks with \\

\input{arxiv/author_info}

\date{\today}

\maketitle
\input{sections/abstract}

%% file: arxiv/author_info.tex
\author{%
{Lim} Ying Chen\\
Faculty of Science\\
National University of Singapore\\
6 Science Drive 2, 117546, Singapore\\
\texttt{e0014951@u.nus.edu}\\
\And
{Rakesh} Das\\
Max Planck Institute for the Physics of Complex Systems\\
Nöthnitzer Str. 38\\
01187 Dresden, Germany\\
\texttt{rdas@pks.mpg.de}\\
\And
{Tetsuya} Hiraiwa\\
Institute of Physics\\
Academia Sinica\\
Taipei 115201, Taiwan\\
\texttt{thiraiwa@as.edu.tw}
\And
N. Duane {Loh}\\
Centre for Bio-imaging Sciences\\
National University of Singapore\\
14 Science Drive 4\\
117557, Singapore\\
\texttt{duaneloh@nus.edu.sg}
}

%% file: sections/abstract.tex
\begin{abstract}
In complex systems, groups of interacting objects may form prevalent and persistent spatiotemporal patterns, which we refer to as motifs.
These motifs can exhibit features that reveal how individual objects interact with one another.
Simultaneously, the motifs can also interact, causing new coarse-grained properties to emerge in the system.

In this paper, we found motifs in a simulated system of Dynamically Self-Organising cells.
We also found that quantifying these motifs with a set of physically interpretable structural and dynamic features efficiently captures the interaction dynamics of the motifs' underlying cells.
Using these motif features, we revealed packing strain and defects in large compact aggregates, semi-periodicity in motif ensembles, and phase space classes with unsupervised machine learning.
Additionally, we trained neural networks to infer the critical hidden microscopic interaction parameters within each motif from coarse-grained motif features extracted from snapshots of the system.
Furthermore, we uncovered emergent features that can predict the movement of cell collectives by hierarchically coarse-graining smaller motifs into larger ones (e.g. motif clusters).
We speculate that this concept of motif hierarchies may be applied broadly to many-body interacting systems that are otherwise too complex to understand.

\end{abstract}

%% file: sections/introduction.tex
\section{Introduction: Motif Hierarchies}
%\subsection{Motif Hierarchies}
There are many interesting examples of complex systems, including the collective behaviour of swarms of insects \cite{vandervaartMechanicalSpectroscopyInsect2019}, schools of fishes \cite{weihsHydromechanicsFishSchooling1973}, and human traffic \cite{nagataniPhysicsTrafficJams2002}.
The study of such complex systems has given us insights about the collective, allowing us to better exploit or improve on their inner workings.

Such complex systems are often abstracted as large numbers of interacting units that behave, as a whole, differently from individual units \cite{newmanResourceLetterCS2011}.
This phenomenon is sometimes known as emergence, which can occur even if the units are only capable of simple interactions.
This emergence is important in many scientific disciplines, like biology and chemistry \cite{newthEmergenceSelfOrganizationChemistry2006}, condensed matter physics \cite{musserEmergenceCondensedMatter2022a} and ecology \cite{schluterCapturingEmergentPhenomena2019}, and its importance is summarised by Philip Anderson's in his 1972 landmark publication `More is Different' \cite{andersonMoreDifferentBroken1972}, with lasting impact \cite{strogatzFiftyYearsMore2022}.

%{\color{red}Complex systems can be difficult to predict when small changes between the interactions of units lead to dramatic changes in their macroscopic behaviour (maybe cite Tetsuya's paper, or paper on chaotic systems).}
%Introduce emergence coming from complex systems
%%As such, this phenomenon of complex systems is key to the complexity behind many fields of research, and its study is an important step to allow for accurate predictions and methods to manage these seemly chaotic systems.

Systems that show emergence sometimes spontaneously form higher level structures when groups of interacting units \cite{dewolfEmergenceSelfOrganisationDifferent2005,dehaanHowEmergenceArises2006,guardiaCategoricalTheoryEmergence2022} display correlated spatiotemporal dynamics. 
When these structures persist in time, they can be viewed as emergent motifs.

%Examples of motifs in nature include: the local tetrahedral molecular motifs in water \cite{wernetStructureFirstCoordination2004}, clusters of biological cells moving together like a transient motif during fetal development of zebrafishes \cite{qianPulsesRhoASignaling2023}, and the chevron-shaped motifs that form in large groups of migratory birds \cite{portugalUpwashExploitationDownwash2014}.

In condensed matter systems, short range motifs are still expected even when long range order is absent (i.e., at high temperatures or entropy).
For example in liquid water, local orientational motifs can appear in liquid water molecules at high temperature \cite{wernetStructureFirstCoordination2004}, which change when supercooled \cite{sellbergUltrafastXrayProbing2014} or subjected to high local electric field \cite{kahkViscousFieldalignedWater2018a}.
Different types of order parameters are used to identify recurrent, persistent atomic motifs especially those that spontaneously arise in three-dimensional molecular dynamic simulations \cite{lazarVoronoiCellAnalysis2022,stukowskiStructureIdentificationMethods2012}.

Relatedly, since most cellular systems lack long-range order, it can be challenging to identify motifs therein.
Various approaches have been used to find and quantify motifs.
These include nematic and hexatic order parameters for structural correlations \cite{grossmannParticlefieldApproachBridges2020,shiSelfPropelledRodsLinking2018,weitzSelfpropelledRodsExhibit2015}, as well as dynamic order parameters for cell motility within a group \cite{zottlEmergentBehaviorActive2016}.
Other methods include grouping adjacent cells using Voronoi tessellations to measure the angular regularity between cell neighbours \cite{dafontouracostaCharacterizingPolygonalityBiological2006}.

%When several interacting units behave as a single motif, their features can be coarse-grained. 
Should we succeed in identifying interacting units as motifs, the features of these units can typically be coarse-grained. 
Such coarse-grained motif features efficiently describe how its member units behave as a group, though at the expense of omitting the motif's internal degrees of freedom.

When motif features can be coarse-grained, groups of correlated motifs can also be meaningfully coarse-grained into higher-level motifs.
Iterating this motif-building process leads to a hierarchy of structural motifs \cite{danLearningMotifsTheir2022}.
In some sense, this hierarchy gives a coarse-grained description of the state and behaviour of a complex many-body system, while retaining more detail than an oversimplified mean-field characterisation.

\begin{figure*}[!ht]
    \centering
	\includegraphics[]{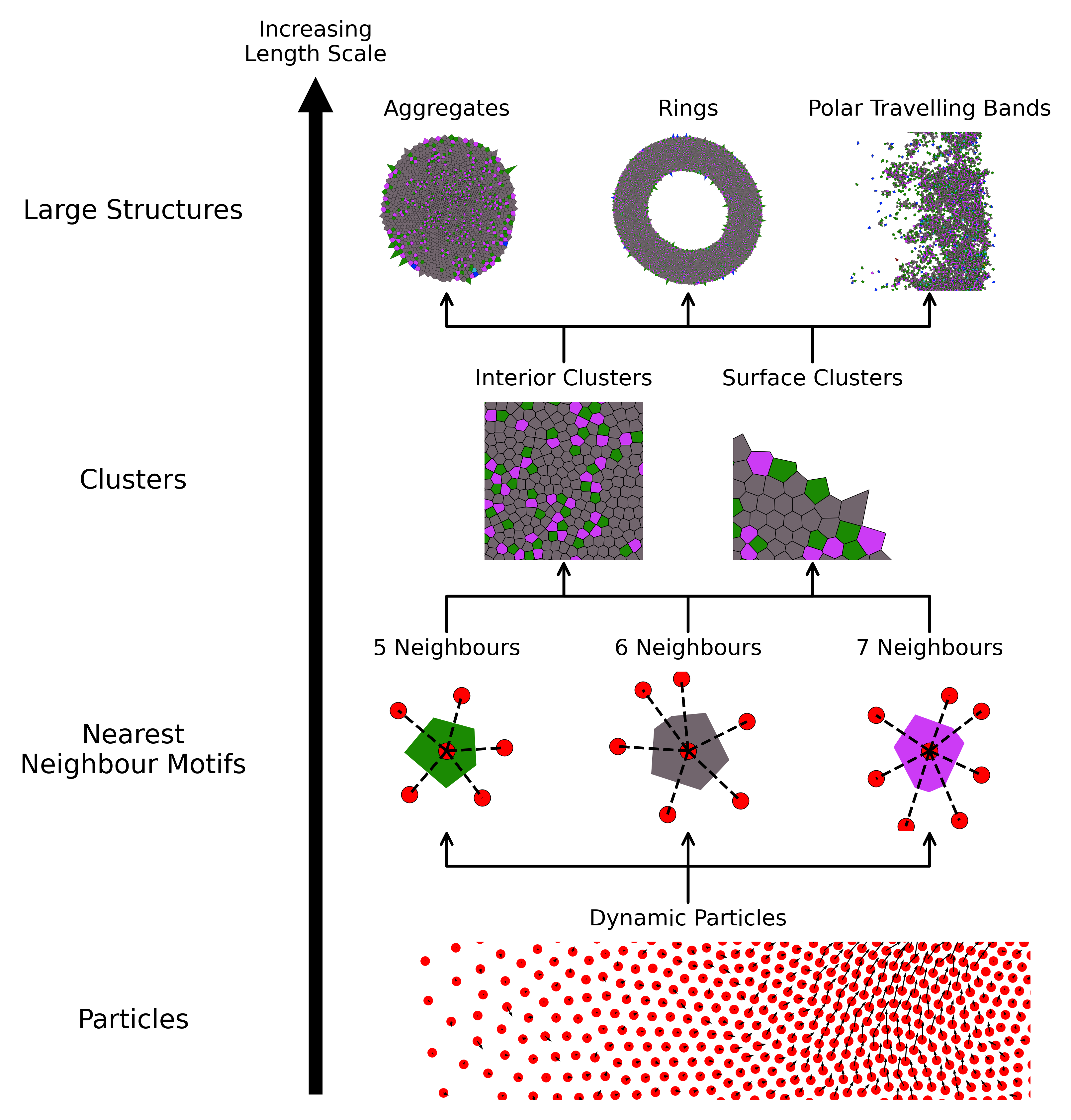}%
	\caption{Hierarchy of motifs formed by Dynamically Self-Organising (DSO) cells \cite{hiraiwaDynamicSelfOrganizationIdealized2020} at increasing length scales. Higher-level motifs can be formed by iteratively grouping low-level motifs together (e.g., grouping particles into nearest neighbour motifs, and nearest neighbour motifs into clusters), with each layer possibly revealing new emergent properties of the system.}%
	\label{fig:length_scale}%
\end{figure*}

%the state of the system is now described by the hierarchy
%mean field does not show locality and interaction between motifs

%Starting from the nearest neighbour motifs, increasingly high level motifs can be iteratively constructed for further coarse graining.
%At each step, we simultaneously coarse-grain groups of units into motifs, and features of individuals in the group into motif features.
%This enables us to give clear descriptions to increasingly large groups of objects, while also simplifying the difficulty with finding the next level grouping.

%Motif hierarchies simplifies the system, making them easier to study
Practically, the behaviour of a complex system can be characterised at any level of the motif hierarchy. 
For example, self-organising groups of cells \cite{hiraiwaDynamicSelfOrganizationIdealized2020} can be understood in four different levels: as a system of dynamic cells, as a collection of $n$-particle groups, as groups cell clusters, or as multiple large aggregate structures (\Cref{fig:length_scale}).
This hierarchy of motifs gives us the flexibility to characterise the system's complex dynamics at multiple length and time scales.

The exercise of building a motif hierarchy from the bottom up can readily reveal emergent descriptions at any level of the hierarchy.
This emergence is clearest when the dynamics of structural motifs at a particular level are absent from the lower motif levels.
%For example, the flocking of birds is a well-known emergent phenomenon, where large groups of birds coordinate to forage for food more efficiently \cite{sridharWhyBirdsParticipate2009}.
%However, it is common for flocks to comprise multiple species, giving hierarchical subgroups that differ in foraging behaviour or substrate \cite{jonesSimilarForagersFlock2020}.
%An example of motif hierarchies comes from social networks \cite{pallaUncoveringOverlappingCommunity2005}.
%Cliques, where each member knows every other member, is a low-level motif that can arise from tight-knit groups of people.
%Strongly overlapping cliques in turn segment the network into communities, which are high-level motifs.
An example where motif hierarchies highlight details of emergence is seen in flocks of American White Pelicans \cite{andersonForagingBehaviorAmerican1991}.
When foraging, smaller groups of pelicans line up in columns, and these pelican columns cooperate to encircle fish.
In this example, the pelicans' column arrangement is a low-level motif, and higher-level emergent motifs more clearly seen as the cooperative interaction between columns.
However, the qualitative nature of these hierarchical descriptions can make them hard to define and uncover.

%Motif hierarchies support a bottom-up understanding of complex systems.
%By systematically learning how motifs at consecutive levels of the hierarchy are related, we break down the study of large complex systems into multiple smaller and simpler steps.
%Simultaneously, we gain knowledge of how cell groups behave at different length scales, enabling cell movements to be predicted or controlled with greater precision.
%This enables for the behaviour of groups of cells, like their movement or stability as an aggregate, to be accurately predicted.
%To study a large complex system, it is often desirable to coarse-grain its many-body dynamics to far fewer degrees of freedom.
%%One path towards coarse-graining is to find motifs within these systems.

%Discuss what we did
To demonstrate that motifs and motif hierarchies can give useful features in complex systems, we apply the concept to simulated movies of dynamically self-organising (DSO) cells \cite{hiraiwaDynamicSelfOrganizationIdealized2020}.
We show how these features can efficiently describe or predict the structural and dynamic properties of large cell collectives.
Additionally, we show that these features can classify and recover the cells' hidden interaction parameters.

\section{Method: Simulated self-organised cells and features that help identify their motifs}
In this paper, we demonstrate that motifs provide descriptive and interpretable features in a simulated model of cells \cite{hiraiwaDynamicSelfOrganizationIdealized2020} that dynamically self-organise (DSO) into a wide variety of patterns (\Cref{fig:dso_phase}).
Each cell interacts only with adjacent cells within a fixed radius, similar to other models that recapitulate flocking behaviours observed in tissue cells \cite{szaboPhaseTransitionCollective2006}, birds \cite{darukaPhenomenologicalModelCollective2009}, insects and fishes \cite{couzinCollectiveMemorySpatial2002,couzinEffectiveLeadershipDecisionmaking2005}.

This gamut of DSO patterns arises from the intercellular interaction representing the effects of cell-cell contact inhibition/attraction of locomotion (CIL/CAL) and contact following (CF).
CIL, which describes how cells tend to separate after contact, was experimentally observed in neural crest cells \cite{scarpaCadherinSwitchEMT2015} and cancer cells \cite{linInterplayChemotaxisContact2015}, while CAL describes the inverse of this phenomenon \cite{li_vivo_2019}.
Relatedly, CF, which describes how contacting cells follow each other, was proposed to explain the behaviour of the slime mold \textit{Dictyostelium Discoideum} \cite{umedaPossibleRoleContact2002}.

Each simulated cell possesses a vectorial polarity that mediates its CIL and CF interactions.
%In the absence of other cells, a cell's polarity determines its velocity.
Specifically, each cell's velocity depends on its polarity and a soft-sphere repulsion against adjacent cells.
This polarity, in turn, is also affected by the polarities of the interacting cells.
For further details, see \cref{SI:model} in the Supplementary Material.

%dso_space

As mentioned above, motifs are correlated structural dynamics formed by groups of cells.
Since these cells interact more strongly when they are close together, we expect such correlations to naturally arise amongst neighbouring cells.
Hence, we compute the Voronoi diagram on the cells' position, and identify motifs by grouping neighbouring cells that share edges in the Voronoi diagram.
Such cell groups are then examined for persistent spatiotemporal features.
When such persistence emerges within a group, we call it a \textit{nearest neighbour motif}.

%From nearest neighbour motifs, we can quantify properties relating to their constituent cells' arrangements and velocity as structural and dynamic features.
The collective properties of groups of cells can be classified into either structural or dynamic features.
Consider the $i^{\text{th}}$ cell, whose nearest neighbours' indices are represented as the set $l_i \equiv \{ \ldots \}$.
Here we focus on the following three structural features that characterise how these neighbours pack around the $i^{\text{th}}$ cell: number of neighbouring cells $n_i$, mean neighbour distance $d_i$, and angular deviation $\sigma_i$;
\begin{subequations}
	\begin{gather}
		n_i=\lvert l_i \rvert \ ,\label{eqn:structural_features:number_neighbours}\\
		d_i=\frac{1}{n_i}\sum_{j \in l_i}\lvert \boldsymbol{r}_i-\boldsymbol{r}_j \rvert \ ,\label{eqn:structural_features:mean_distance}\\
		\sigma_i = \left[\frac{1}{n_i} \sum_{j=1}^{n_i} \left[\theta_{i,j} - \left<\theta_{i,k}\right>_k\right]^2\right]^{1/2} \label{eqn:structural_features:angular_deviation}\;.
	\end{gather}
    \label{eqn:structural_features}
\end{subequations}
We denote the velocity and the speed of the $i^{\text{th}}$ cell as $\boldsymbol{v}_i$ and $v_i$, respectively.
Here we examine the following dynamic features of this cell's neighbours (i.e., $l_i$): their mean dot product $\alpha_i$, divergence $\delta_i$, and curl $\gamma_i$:
\begin{subequations}
	\begin{gather}
		v_i = \lvert \boldsymbol{v}_i \rvert \ ,\\
		\alpha_i = \frac{1}{n_i}\sum_{j \in l_i} \boldsymbol{v}_i \cdot \boldsymbol{v}_j \ ,\\
		\delta_i = \frac{1}{n_i} \sum_{j \in l_i} \frac{\left(\boldsymbol{v}_j - \boldsymbol{v}_i\right)\cdot \left(\boldsymbol{r}_j - \boldsymbol{r}_i \right)}{\lvert \boldsymbol{r}_j - \boldsymbol{r}_i \rvert} \ ,\\
		\gamma_i = \frac{1}{n_i} \sum_{j \in l_i} \frac{\left( \boldsymbol{r}_j - \boldsymbol{r}_i \right) \wedge \left(\boldsymbol{v}_j - \boldsymbol{v}_i\right) }{\lvert \boldsymbol{r}_j - \boldsymbol{r}_i \rvert} \ .
        \label{eqn:curl}
	\end{gather}
    \label{eqn:dynamic_features}
\end{subequations}

In what follows, we use the above structural and dynamic features to capture the correlations between the $i^{\text{th}}$ cell with its $l_i$ neighbours.
We found such features gave descriptive quantities that characterised the interactions between DSO cells, enabling us to study emergent structural and dynamic behaviours of cell collectives.
Additionally, we found that these quantitative features provided ideal features for machine learning, enabling us to classify `phases' on the parameter space of interaction forces and also to regress on the interaction strength of cells from their trajectories.

\begin{figure*}
    \centering
	\subfloat[Cell neighbourhoods colourised by structural motifs features from \cref{eqn:structural_features}.]{%
		\includegraphics[]{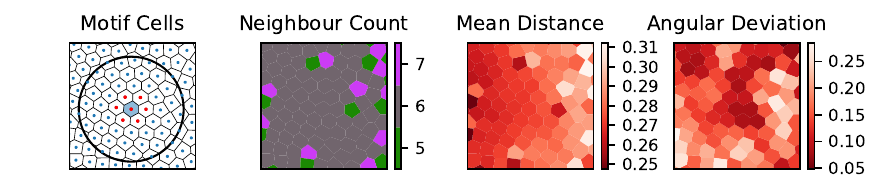}%
        \label{fig:structural_motifs}%
	}\\
	\subfloat[Cell neighbourhoods colourised by dynamic motifs features from \cref{eqn:dynamic_features}.]{%
		\includegraphics[]{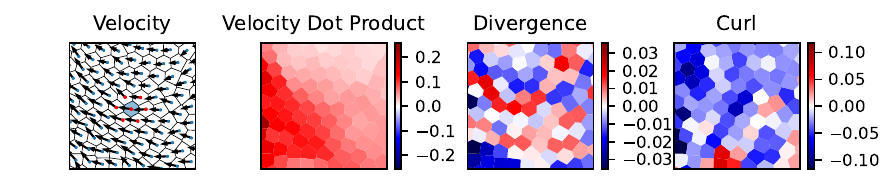}%
        \label{fig:dynamic_motifs}%
	}%
	\caption{Quantifying local structural and dynamic properties with motif features.}%
	\label{fig:motifs_features}%
\end{figure*}

%% file: sections/meaningfulness.tex
\section{Evidence of meaningful motifs}
\subsection{Establishing the persistence and prevalence of nearest neighbour motifs}
We surmise that meaningful structural motifs should have prevalent and persistent cell group features.
A prevalent feature in the model under study \cite{hiraiwaDynamicSelfOrganizationIdealized2020} is that cell groups often have six nearest neighbours (6NNs).

\begin{figure*}[!t]
    \centering
    \subfloat[Prevalence of nearest neighbour cell groups, showing that 6NNs are prevalent under most conditions.]{%
    \includegraphics[]{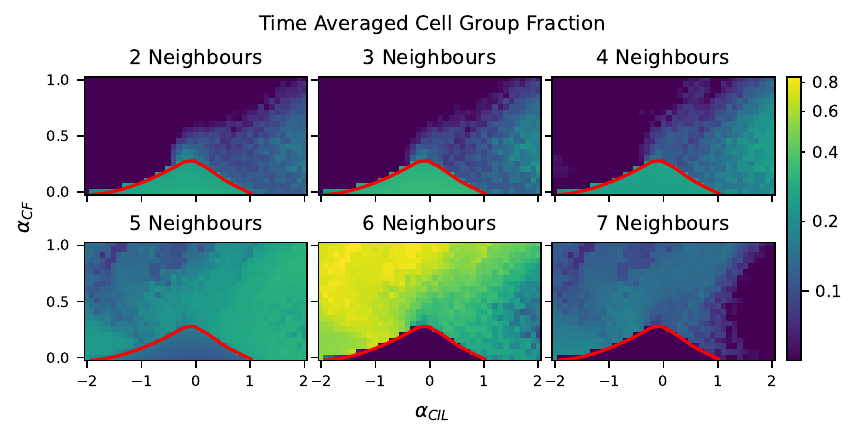}%
    \label{fig:nn_prevalence}}\\
    \subfloat[Persistence of cell groups with various nearest neighbours counts. Groups with 6NN and 4NN had especially long lifetime when $\alpha_{\text{CIL}} < 0$.]{%
    \includegraphics[]{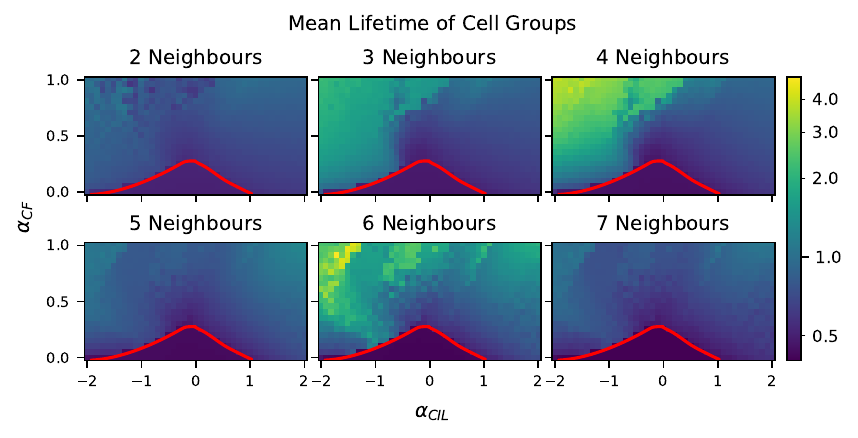}%
    \label{fig:nn_persistence}}%
    {\hypersetup{linkcolor=black}\caption{Prevalence and persistence of different nearest neighbour cell groups at various contact following ($\alpha_{\text{CF}}$) and contact inhibition of locomotion ($\alpha_{\text{CIL}}$) strengths. A red line is drawn to highlight the region with disordered dynamic self-organising (DSO) cell pattern.}}%
    \label{fig:prevalence_and_persistence}%
\end{figure*}

Cell groups with six nearest neighbours (6NNs) are prevalent in the non-disordered region of the parameter space (see \Cref{fig:dso_phase}).
When averaged over time, such groups had high fractional prevalence, meaning they account for a large fraction of nearest neighbour cell groups.
This prevalence is clearest at $\alpha_{\text{CIL}} < 0$, where large quasi-stable and compact aggregates are formed.
Since 6NNs also prevailed in close-packing of hard \cite{katgertJammingGeometryTwodimensional2010} and soft \cite{irannezhadPackingSoftSpheres2023} spheres, its prevalence here is unsurprising because our simulation includes soft-sphere repulsion.
As we shall see in \ref{packing}, dense clusters of 6NNs are often accompanied by 5NN and 7NN defects, which explains their high occurrence in \cref{fig:nn_prevalence}.

Correspondingly, 6NNs are rare in the disordered region.
Here, weak intercellular interactions (CIL and CF) result in highly dispersed cells where 2-4NNs dominate instead.
Similarly, 4-5NNs are preferred at $\alpha_{\text{CIL}} > 0$, where cells are relatively dispersed due to their repulsion.

Candidate motifs should also persist in time.
When there is sufficient interaction strength, \cref{fig:nn_persistence} shows that the prevalent 6NN cell groups are also persistent.
In general, the mean lifetime of motifs increased with larger $\alpha_{\text{CF}}$, ranging from 0.8 in static aggregates up to 4.0 in rotating aggregates.

We note in passing that while 4NN cell groups do not appear prevalent in \Cref{fig:nn_prevalence}, they often comprise around 50\% of surface cell groups on aggregates (\Cref{fig:aggregate_structural_features}).
Such groups typically move slowly along the aggregate's surface and persistent like 6NNs.
Compared to cell groups in the interior of aggregates, these 4NN surface groups tend to have high angular deviation ($\sigma>0.5$) because their neighbours are concentrated towards the aggregate's interior.

\subsection{Detecting dynamical strain and defects in cell collectives with motif structural features}
%\subsection{Dynamically changing strain patterns and defects in motif aggregates}
\label{packing}

\begin{figure*}[!t]
    \centering
    \includegraphics[]{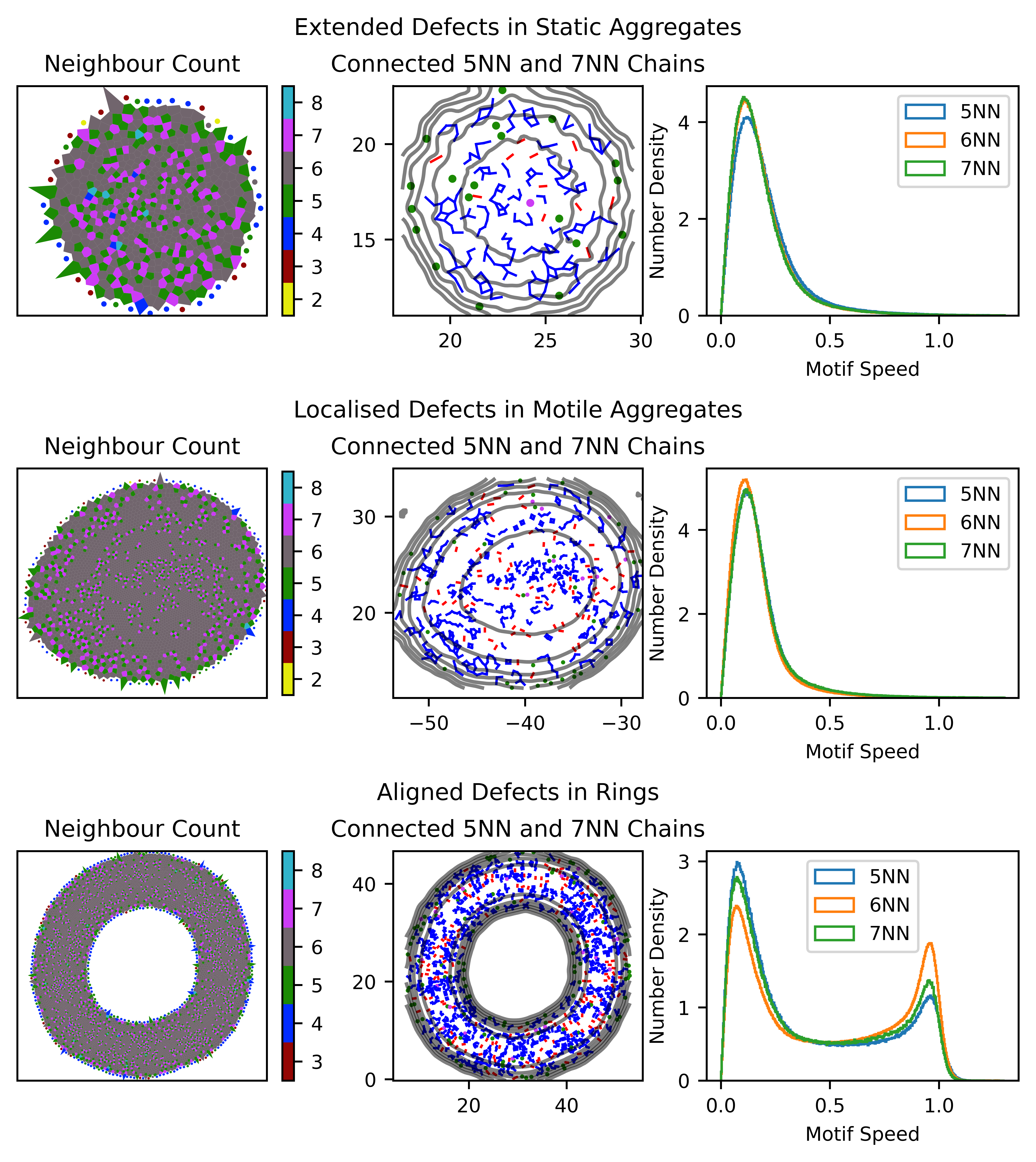}%
    \caption{(LEFT) Voronoi cells coloured by their number of neighbours, showing the prevalence of 6 nearest neighbours (6NN) groups in static aggregates, motile aggregates and rings. (CENTRE) Neighbour distance contour lines (gray) showing strained and increasingly dense packing towards the centre of aggregates. Packing defects are evident from the isolated, adjacent pairs of 5/7NN cell groups (red segments), as well as the extended chains of alternating 5/7NN groups (blue lines). Adjacent defect pairs in rings are also generally radially aligned. (RIGHT) Distribution in speed of cell groups shows a second peak near 1.0 when cells circulate particularly quickly, like in rings.}%
    \label{fig:aggregate_structural_features}%
\end{figure*}

%Motifs do not have to be static.
%Aggregates (i.e., static, motile, or rotating aggregates, when $\alpha_{\text{CIL}} < 0$) are primarily comprised of 6NN motifs (\Cref{fig:aggregate_structural_features}). 
%These aggregates persist when they exceed a critical number of motifs, and when their motifs have inward-pointing polarities (\Cref{fig:PersistentAggregate})).
%Cell groups, which as we show below, are often unstable in isolation but can form persistent motifs with diverse features when clustered.
Rings and aggregates (i.e., static, motile, and rotating aggregates) are mainly composed of persistent 6NN motifs interspersed with defects comprising 5NN and 7NN motifs (\Cref{fig:aggregate_structural_features}).
These structures are typically formed when cells follow or attract strongly, pushing the cells into 6NN motif packings.
Although the 6NN motifs resemble those found in hexagonal nets in materials, they show several structural differences.
%Despite being mostly composed of just these three types of motifs, the interaction between the motifs has produced three interesting structural variations and properties.
%and temporal variations.

Unlike hexagonal crystals, motifs in this system are highly compressible.
This compressibility shows up as strain patterns that are unusual for solid-state materials.
For example, strain in aggregates presents itself as an increasingly dense packing of 6NNs deep within the aggregate.
Because aggregates have inward-pointing polarities (\Cref{fig:aggregate_polarity}), their cells tend to collapse toward their centre.
A catastrophic inward collapse due to the polarity is prevented by the net outwards pressure created by the differential packing density of the aggregate.

Motifs near the surface of aggregates also tend to be more irregular.
Such motifs have larger standard deviations in the bond angle between adjacent neighbours (i.e., angular deviation) since they generally consist of cells with neighbours on only one side (see \cref{fig:dense_sparse_cutoff}).
Relatedly, the angular deviation of surface motifs tends to increase rapidly along with their mean neighbour distance, likely because the weaker inter-cellular forces at long distances let cells move more freely.
In contrast, the angular deviation of motifs deeper within aggregates hovers around $\sim 0.17$ rad regardless of the motifs' mean neighbour distance.

%2
Second, 5NN and 7NN motifs in aggregates and rings often resemble dynamically fluctuating and extended defects that can move (\Cref{fig:aggregate_structural_features}).
This is unlike many solid-state systems, where such packing defects are stationary and prevents the underlying objects from moving.
%Second, the 6NN motifs are interspersed by a wide range of prevalent 5NN and 7NN cell groups that resemble dynamically fluctuating defects that may be taken as higher-level motifs (\Cref{fig:aggregate_structural_features}).
While a number of these purported defects were isolated 5/7NN groups (shown as points in the figure), many instead form neighbouring pairs (shown as red lines), and even more form long alternating 5/7NN chains (blue lines).
This prevalence of nearest-neighbour motif pairs and chains hints that the defects may be higher-level motifs.
%, although they are short-lived relative to 6NN motifs, as seen in \cref{fig:nn_persistence}.

Although these 5NN and 7NN packing defects can move, the defects tend to have a stable probability density (\Cref{fig:aggregate_structural_features}).
The defect density within such cell collectives is often correlated to how motifs circulate within the collective.
Such correlation can be seen in motile aggregates, where defects generally gather near the aggregate's surface or alongside the aggregate's central fast-flowing stream.
In some cases, these defects can also become aligned, as seen by the radially aligned 5/7NN defect pairs in rings.

We note that the speed distribution of cell groups (both 6NN and 5NN/7NN defects) can show multiple peaks when cell collectives have very fast-moving internal circulations.
In such cases, the cell collective tends to be elongated, with many of the motifs in the fast-moving stream reaching a peak speed near 1.0.

Third, the 6NN motifs in aggregates become stable only in sufficiently large numbers (see \cref{fig:unstable_comoving_hex,fig:PersistentAggregate}).
This is perhaps similar to the critical nucleus size in classical nucleation theory.
The difference here is that there no concept of energetics in these cellular aggregates.
Locally, the cells of each 6NN motif within the aggregate tend to have co-aligned polarities (\Cref{fig:aggregate_polarity}).
These co-polarity 6NN motifs are unstable if extracted and isolated from their original aggregate (\Cref{fig:unstable_comoving_hex}).
%This result is consistent with an earlier stability study (cite Tetsuya's paper): a 6NN motif can be formed at the intersection of three linear chains that are $60^{\circ}$ rotated from each other, and only one of these three chains is stable (see SI).
Although the 6NN motif are individually unstable, they gain a surprising collective stability when sufficiently many of them are packed into an aggregate with a radially inward polarity field (\Cref{fig:PersistentAggregate}).
Relatedly, the central motifs within aggregates are not stationary.
Cells there either buoy towards the aggregate's surface (i.e., static aggregates) or move in a stream (i.e., motile and rotating aggregate).

%unstable_comoving_hex

%Finally, the pair-wise polarity-mediated interactions between the cells in a 6NN motif grow weaker as the motif expands.
%Since the cells in a 6NN motif tend to be arranged as a hexagon, all pairs of cells would experience mutual forces when the mean distance between the central cell and its nearest neighbours $d<0.5$. 
%Conversely, when $d>0.5$ the distance between extremal cells starts to exceed unity.
%As such, these pairs of extremal cells can only interact indirectly through the forces mediated by the central cell.

%dense_sparse_cutoff

\subsection{Correlation between aggregate dynamics, its central cells, and motif lifetime}
%\subsection{Aggregate dynamics and motif lifetimes are correlated with its central cells}
Based on how the central cells within an aggregate move, the behaviour of the aggregate and the lifetime of motifs within it can be inferred.
As previously mentioned, motifs at the centre of aggregates tend to have shorter lifetime and cells there are continually pushed outwards.
For aggregates to persist without breaking apart, these ejected cells must be recaptured, which requires that the aggregate be sufficiently large.
Depending on the intercellular contact following (CF) strength, the central cells may be ejected in a random or aligned direction, giving different aggregate behaviours.

Despite static aggregates appearing stationary as a whole, they typically have short-lived motifs.
The central cells within these aggregates buoy outwards in random directions due to weak contact following forces (i.e., small $\alpha_{\text{CF}}$).
This scattered movement disrupts other motifs and reduces their lifetime.
The motif lifetime of static aggregates are further shortened as the cells constantly jiggle about due to noise dispersion.
This motif instability appears as a markedly shorter motif lifetime in static aggregates compared with motile and rotating aggregates (see \cref{fig:prevalence_and_persistence}).
%(\Cref{fig:motif_lifetime})

\begin{figure*}
    \centering
    \includegraphics[]{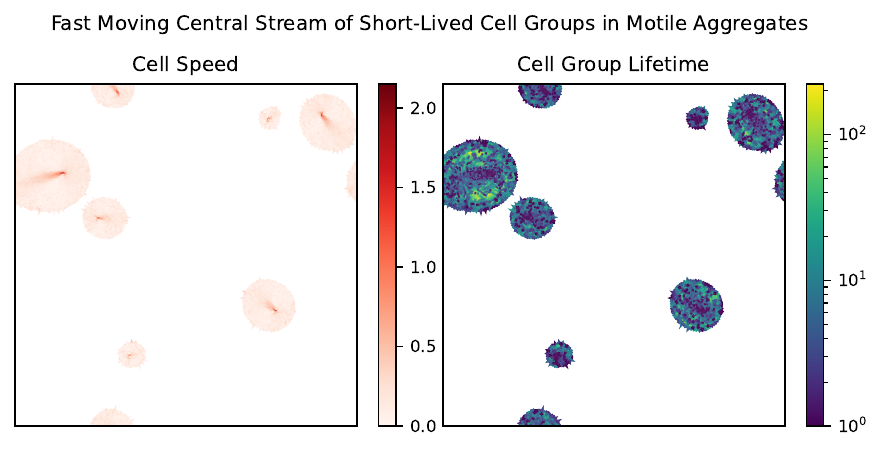}%
    \caption{Cell groups in aggregates can have significantly longer lifetime when located outside of the aggregate's central fast moving stream.}%
    \label{fig:aggregate_dynamic_features}%
\end{figure*}

Motile and rotating aggregates, which form at larger $\alpha_{\text{CF}}$, can move because they have a stable internal circulation of cells \cite{hiraiwaDynamicSelfOrganizationIdealized2020}.
At the aggregate's centre, the ejected cells become aligned and form a central fast-flowing stream (\Cref{fig:aggregate_dynamic_features}) due to the strong contact following forces (i.e., large $\alpha_{\text{CF}}$).
These cells are brought to the aggregate's front before being slowly pushed back to the rear by moving along the sides.
As a whole, this circulation creates a net centre of mass movement without any loss in cells.

%aggregate_dynamic_features

Within these motile and rotating aggregates, 6NN motifs have a longer mean lifetime.
Cells located outside of the central stream tend to move more orderly and slowly (\Cref{fig:aggregate_dynamic_features}), which correlated with higher motif stability.
In particular, exceptionally long-lived motifs with lifetimes ten times longer than the aggregate's average can form along the side of the central stream.
The likelihood of forming these long-lived motifs is correlated with the average motif's lifetime, with them appearing most often in rotating aggregates.
%Large aggregates were also observed to better support long-lived cell groups.

\subsection{Detectable temporal semi-periodicities of motif ensembles}
%\subsection{Fingerprinting stability and periodicity of multicellular structures using localised features}
When cells spontaneously form large-scale dynamic self-organised (DSO) patterns, they also display recognisable local features.
%there is a corresponding change in the local scale cell packing structure and interaction forces.
Specifically, we expect the structural features of nearest neighbour motifs to change along with the evolution of large multicellular structures (e.g., aggregates, rings, spirals, etc.).
Hence, these local motif features should fingerprint the behaviour of these larger structures.

\begin{figure*}
    \centering
	\includegraphics[]{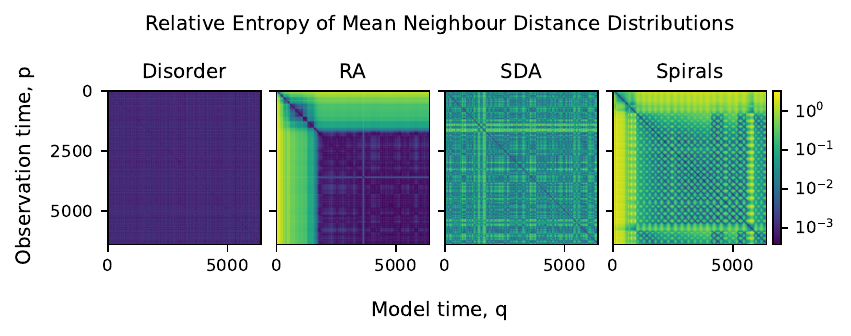}%
	\caption{Mean neighbour distance of motifs revealing systems evolving into a stable equilibrium (Rotating Aggregates), unstable equilibrium (Snakelike Dynamic Assembly), or cyclic state (Spirals).
    Relative entropy is used to measure the degree at which a system has changed via a pair-wise comparison of neighbour distance distribution between each pair of frames as the system evolves ($D_{\text{KL}} = \sum_{x\in \left\{y: q(y) > 0\right\}} p(x) \ln{\frac{p(x)}{q(x)}}$).}%
	\label{fig:distribution_time_correlation}%
\end{figure*}

DSO cells evolve differently based on their interaction parameters, reaching a stable state, having unstable patterns, or periodically cycling between modes.
Here, we characterise the local dynamics at each simulation time using the mean neighbour distance of motifs 
This quantity encodes both the local cell packing density and distance-dependent interaction forces acting on each cell.
%To simplify the computation, the distributions were calculated from cell groups without regard to their number of neighbours.
For each simulation time, there is a distribution of mean nearest neighbour distances.
Rather than describing this distribution by just its mean and variance, we quantise (i.e., coarse-grained) these features into a histogram.
We then compare pairs of such histograms at different times using their relative entropy. 
\Cref{fig:distribution_time_correlation} shows this relative entropy between all pairs of time points as a system evolves, with dissimilar distributions having high entropy.

Disordered DSO patterns tend to be time-invariant.
This is evident from its low relative entropies in \cref{fig:distribution_time_correlation}.
%show little change in their [distance distributions] over time.
Here, the initially scattered cells remain scattered, as their pair-wise interactions are too weak for them to organise into persistent patterns.

With certain combinations of strong contact following (large $\alpha_{\text{CF}}$) and contact attraction of locomotion ($\alpha_{\text{CIL}}<0$), cells can spontaneously organise into stable structures like aggregates (e.g., rotating aggregates in \cref{fig:distribution_time_correlation}).
The internal cell dynamics of aggregates are more stable and ordered compared to structures from other DSO patterns.
This is evident in \cref{fig:distribution_time_correlation} from the reduction in relative entropy after cell structures were formed.
%Occasionally, simulations that were burned in can still contain multiple aggregates.
%Such systems would appear stable for some time, but show moderate changes in their distribution when these aggregates merge.

When cells show contact inhibition ($\alpha_{\text{CIL}}>0$), their local features fluctuate rapidly.
Hence, the large-scale structures found are generally unstable, with patches breaking apart and joining repeatedly.
This instability correlates with the fluctuating neighbour distance distributions, which appear as recurrent time points with high relative entropy.

Finally, the local structure of cells periodically cycle between two modes when the cells self-organise into spirals.
These cell collectives alternate between an elongated and a rotund mound, where the elongated mound turns back into itself and evolves into its rotund form.
This is clearly evident in \cref{fig:spiral_periodic}, whose relative entropies show a periodic alternation between these two forms.

%spiral_pattern_periodic

%% file: sections/results.tex
\section{Motifs have ideal features for machine learning}
As we have seen in \cref{fig:aggregate_dynamic_features}, cell dynamics can be correlated to the motif features of the cells.
This correlation, in reverse, suggests that motif features can predict such dynamics.
Even so, finding the correct and compact set of features that can be used for predictions is non-trivial especially given the large number of cells and their changing dynamics over time.
%For example, if we were to watch a movie of a collection of cells, which features should we extract to predict their behavioural phase (\cref{phase_space}), their most likely motion of cell groups (\cref{predict_movement}), and the cells interaction strengths (\cref{predict_parameter})?

In this section, we show that useful motif features for studying a system can come from anywhere in the motif hierarchy in \cref{fig:length_scale}.
By extracting low-level motifs from nearest neighbour cell groups, we can predict the interaction parameters $\alpha_{\text{CIL}}$ and $\alpha_{\text{CF}}$ associated with each system, and also the behavioural phases from varying these parameters.
Additionally, we show that high-level motif features from cell clusters can predict the movement and rotation of large multicellular rotating aggregates.

\subsection{Unsupervised determination of DSO phase space}
\label{phase_space}
By visually inspecting the dynamics and density for each of the 861 unique pairs of interaction force parameters $\alpha_{\text{CIL}}$ and $\alpha_{\text{CF}}$, the author of \cite{hiraiwaDynamicSelfOrganizationIdealized2020} determined a `phase diagram' of its types of dynamics.
Here, we explore if such phases can be obtained with little human supervision.

To uncover this `phase diagram', we quantified the local neighbourhood structural and dynamic features around each cell (see \cref{fig:motifs_features}).
These six features of the $n^{\text{th}}$ cell at time $t$, $\featAfull{}$, are listed in \cref{fig:movie_phase_cluster}.
The many motif features from each movie were reduced to a single 72-dimensional feature vector $\featCfull{}$: first by coarse-graining $\featAfull{}$ over all the $n$-cells at each time $t$ to obtain $\featBfull{}$, then again over all $t$ to $\featCfull{}$.
We then embedded these 72-dimensional feature vectors $\featCfull{}$ into a two-dimensional space in \cref{fig:movie_phase_cluster}, attempting to place movies with similar features close together using Uniform Manifold Approximation and Projection (UMAP).

\input{sections/phase_classification_figure}

\Cref{fig:movie_phase_cluster} shows that the UMAP embedding of $\featCfull{}$ forms well-separated groups.
We colourised these points by their respective cluster labels found through Hierarchical Density-Based Spatial Clustering of Applications with Noise (HDBSCAN), as well as their corresponding $\alpha_{\text{CIL}}$ and $\alpha_{\text{CF}}$ interaction space in \cref{fig:movie_phase_cluster}.

The HDBSCAN clusters in \cref{fig:movie_phase_cluster} mostly overlap with the `phases' identified via visual inspection in \cite{hiraiwaDynamicSelfOrganizationIdealized2020} shown in black.
Presumably, the clearest overlap was in the disordered `phase', whose distinctive motif features (e.g., low 6NN prevalence as seen in \cref{fig:prevalence_and_persistence}) caused its embedded points to form an exceptionally distinct cluster.
However, our clustering differs from the visually identified `phases' in \cite{hiraiwaDynamicSelfOrganizationIdealized2020} in some regions (e.g., polar traveling bands vs. homogeneous polar flocks).

While \cref{fig:movie_phase_cluster} shows that the coarse-grained feature vectors of each movie, $\featCfull{}$, can automatically identify `phases', it can be difficult to experimentally measure the requisite $\featAfull{}$ features for many $n$ cells, each over long periods $t$.
Here, we explore the variations in `phases' obtained when we have incomplete information: either with limited time samples or with a limited number of cells.

Consider time-limited observations, such as when consecutive time snapshots of a system are available.
We reproduce such a scenario by extracting feature vectors \cref{fig:frame_phase_cluster} at only 40 well-separated movie snapshots from each of the 861 movies (i.e., $\vec{A}(n,t \in \{t_1, t_2,\ldots, t_{40}\})$).
These feature vectors were coarse-grained over space to obtain $861\times 40 = 34,440$ feature vectors $\featBfull{}$ for embedding with UMAP.
Again, the embedding points were clustered and labelled using HDBSCAN, and each $\alpha_{\text{CIL}}$ and $\alpha_{\text{CF}}$ parameter pair was colourised by the most common cluster label obtained from its 40 movie snapshots in \cref{fig:frame_phase_cluster}.

`Phases' can also be detected by only tracking the dynamically changing features around individual motifs over long periods.
From each of the 861 movies above, we extracted feature vectors around the 100 cells that most often remained 6NN motifs (\Cref{fig:group_phase_cluster}).
The feature vectors of each of these 100 cells were accumulated for over 400 time points, resulting in $861 \times 100 \times 400 = 34,440,000$ feature vectors $\featAfull{}$.
These feature vectors were temporally coarse-grained as 86,100 feature vectors $\featDfull{}$, one for each of the 100 cells interacting with its own $\alpha_{\text{CIL}}$ and $\alpha_{\text{CF}}$ parameters. 
These $\featDfull{}$ feature vectors were then embedded again using UMAP and clustered with HDBSCAN to give \cref{fig:group_phase_cluster}.
Here, each pair of $\alpha_{\text{CIL}}$ and $\alpha_{\text{CF}}$ parameters was coloured by the most common cluster label of its 100 tracked cells.

The `phase' boundaries from differently coarse-grained features (i.e., \cref{fig:phase_cluster}) showed noticeable differences.
Such differences are perhaps unsurprising since each set of coarse-grained features only contains a partial perspective of the spatiotemporally changing features in the complex DSO interactions: either the spatial (\cref{fig:frame_phase_cluster}) or temporal (\cref{fig:group_phase_cluster}) details were coarse-grained away, or that both sets of features were heavily coarse-graining (\cref{fig:movie_phase_cluster}).

Yet the disordered phase emerged consistently amongst all three clusterings (i.e., \cref{fig:phase_cluster}) and also that from visual inspection (guided by cell density).
This consistency indicates that the disordered phase can be robustly detected despite spatial or temporal coarse-graining.
Nevertheless, the `phase' boundaries between rings, spirals, aggregates, bands, and flocks are more subjective since they depend on the type of coarse-grained feature that was examined.

%Supposing that the `phases' uncovered from low-level motifs (i.e., nearest neighbour cell group) are emergent properties of the model, higher-level motifs may also give useful features for modelling other emergent phenomena.
%In \cref{predict_movement}, we find that small clusters of cell groups can give features that predict the movement of cell aggregates.

%One could presuppose that emergence manifests through decision boundaries in some high dimensional feature space (e.g., between disorder and aggregates in \cref{fig:frame_phase_cluster}).
%Such decision boundaries clearly appear all the down to local features of individual 6NN motifs.
%Hence, one may expect aggregated features that more efficiently describe the behaviour of these emergent higher-level motifs (e.g., aggregates).

\subsection{Emergent vortices can predict an aggregate's motion}
\label{predict_movement}
\begin{figure*}
    \centering
    \includegraphics[]{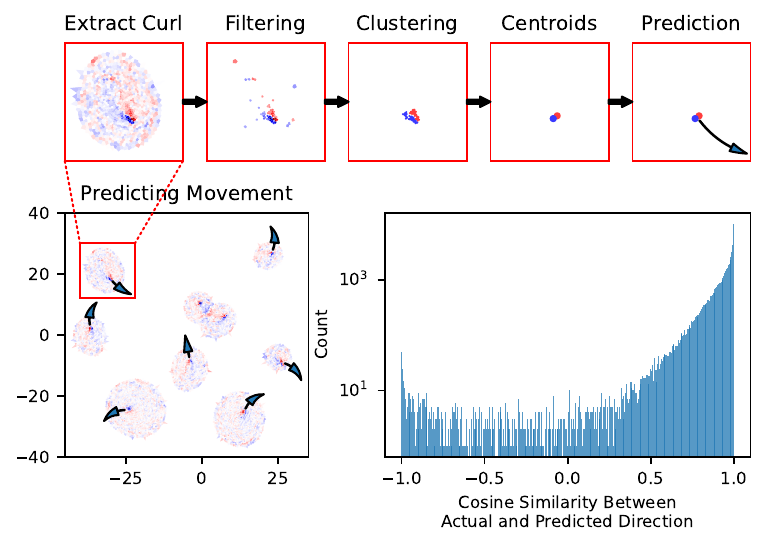}%
	\caption{Coarse-grained dynamic features for predicting the movement and rotation of rotating aggregates. Negative and positive motif curl groups are coloured blue and red respectively. BOTTOM LEFT: cell groups coloured by their curl (defined in \Cref{eqn:curl}). TOP ROW: The curls of an aggregate's cell groups are coarse-grained to locate the aggregate's dipolar vortices. LOWER RIGHT: Cosine similarity between each aggregate's predicted and actual centre of mass direction shows good alignment (i.e., $+1$ similarity) with an exponential falloff.}%
	\label{fig:movement_prediction}%
\end{figure*}

Recall that \cref{fig:aggregate_dynamic_features} showed how the internal cellular circulation of rotating aggregates is correlated with the aggregates' movement.
%With rotating aggregates, this cellular circulation also resulted in a stable clockwise or anti-clockwise rotation of the aggregate.
%Here, we show that vortices can emerge from motifs that, in turn, help us predict the movement and rotations of rotating aggregates.
Often, this circulation causes a pair of persistent dipolar vortices to form near the aggregate's front (see high-curl regions in the aggregate in the top panel of \cref{fig:movement_prediction}).
Here, we extract motif features from the cell clusters making up these persistent vortices and use these features to predict the movement and rotations of rotating aggregates.

%within each aggregate, cell groups are allowed to merge with others if they have speed in 97%tile, curl below 3%tile or above 97%tile
%cell groups can only merge with those adjacent to it can have the same curl sign (positive/negative)
%driving clusters must have at least 5 cel groups
The internal cellular circulation of rotating aggregates tends to correlate with the formation of a dipolar vortex near the aggregate's front (see high-curl regions in the aggregate in the top panel of \cref{fig:movement_prediction}).
The locations of these dipolar vortices can be extracted using the following coarse-graining recipe.
We locate high-curl cell groups using a two-tailed outlier criterion: we only keep cell groups whose curls are in the top or bottom three percentile within each aggregate.
%These high-curl cell groups of the same polarity were spatially clustered together.
Adjacent high-curl cell groups of the same polarity were then clustered together.
Afterwards, tiny clusters containing fewer than five cell groups were excluded.
The surviving clusters with the maximum and minimum total curl were identified as the positive and negative vortices, respectively.

The polarities of each rotating aggregate's dipolar vortices are correlated with the direction in which its cells circulate (see \cref{fig:aggregate_cell_velocity}).
Because the cells in the central stream of this circulation move faster than those that loop back around the aggregate's periphery, the aggregate's centre of mass tends to move in the direction of central circulation. 
Hence, the polarities and positions of an aggregate's dipolar vortices can be used to predict the aggregate's direction of travel.

%aggregate_cell_velocity

Additionally, the relative strengths of the dipolar vortices within each rotating aggregate can accurately predict the latter's turning direction (lower right panel of \cref{fig:movement_prediction}).
Aggregates tended to rotate clockwise when the total curl in the positive curl cluster had a larger magnitude than the negative curl cluster and vice versa.

%A caveat of the procedure is that the predicted movements is invalid when multiple aggregates graze without fully merging.
%Individual aggregates within such collections can be difficult to identify, but they can be detected by the presence of multiple fast-flowing streams within them.
%One way to identify these grazing aggregates is through a similar clustering exercise on the speed of NN motifs.
%In each aggregate, adjacent NN motifs with speed in the top 3\% are clustered, and the number of clusters with at least 5 members counted.
%If at more than one such clusters are found, the aggregate is assumed to formed by grazing aggregates.

We verified the accuracy of these predictions on the trajectories of rotating aggregates with interaction strengths $-2.0 \leq \alpha_{\text{CIL}} \leq -1.5$ and $0.8 \leq \alpha_{\text{CF}} \leq 1.0$.
The cosine similarity between the predicted and actual velocity of each aggregate's centre of mass \cref{fig:movement_prediction} showed exponentially good alignment.
Similarly, \cref{fig:aggregate_turning} showed strong correlations between the aggregate's turning direction and the relative magnitude between its dipolar vortices.

%aggregate_turn

\subsection{Inferring CF and CIL strengths from motif features}
\label{predict_parameter}
When attempting to mathematically model experimental systems, the chosen model and model parameters should accurately reproduce the dynamics seen in the experiments.
Motif features, which were shown to characterise the dynamics of DSO cells, may be used to create quantitative metrics to compare between models and find suitable parameters for them.

In other words, motif features extracted from movies of simulated models alone should encode information that allow the model's parameters to be recovered.
Here, we show that neural networks can recover the $\alpha_{\text{CIL}}$ and $\alpha_{\text{CF}}$ parameters of this DSO cell model when passed motif features coarse-grained over all cells found at a single time point.
Compared to the earlier embeddings of just the raw motif features, these neural network also fold in information about the interaction strengths.
Hence, the embedding space learned by these neural networks are more revealing than the embedding learned by the raw motif features.

\begin{figure*}
\subfloat[Structural and dynamic cell group features of the $n^{\mathrm{th}}$ cell at time $t$, $\featEfull{}$, coarse-grained over all cells at each time in six different ways to give a 42-dimensional feature vector $\featFfull{}$.]{%
\begin{minipage}{\textwidth}
\footnotesize
\centering
\begin{NiceTabular}[t]{|c|p{3em}c}
\hhline{-~-}
$\featEfull{} \in \mathbb{R}^{7}$ & \multicolumn{1}{c|}{} & \multicolumn{1}{c|}{$\featFfull{} \in \mathbb{R}^{42}$}\\\hhline{=~|=|}
Neighbour Count      & \multicolumn{1}{c|}{} & \multicolumn{1}{c|}{$\underset{n}{\operatorname{min}}{\, \featEfull{}}$}\\\hhline{-~-}
Mean Distance        & \multicolumn{1}{c|}{} & \multicolumn{1}{c|}{$\underset{n}{\operatorname{max}}{\, \featEfull{}}$}\\\hhline{-~-}
Angular Deviation    & \multicolumn{1}{c|}{} & \multicolumn{1}{c|}{$\underset{n}{\operatorname{mean}}{\, \featEfull{}}$}\\\hhline{-~-}
Centre Speed         & \multicolumn{1}{c|}{} & \multicolumn{1}{c|}{$\underset{n}{\operatorname{var}}{\, \featEfull{}}$}\\\hhline{-~-}
Velocity Dot Product & \multicolumn{1}{c|}{} & \multicolumn{1}{c|}{$\underset{n}{\operatorname{skew}}{\, \featEfull{}}$}\\\hhline{-~-}
Divergence           & \multicolumn{1}{c|}{} & \multicolumn{1}{c|}{$\underset{n}{\operatorname{kurt}}{\, \featEfull{}}$}\\\hhline{-~-}
Curl                 &&\\\hhline{-~~}
\CodeAfter
\begin{tikzpicture}
    \foreach \x in {2.5,...,8.5}{
        \draw (\x-|2) -- (2.5-|3);
        \foreach \y in {3.5,...,7.5}{
            \draw[color=black!25] (\x-|2) -- (\y-|3);
        }
    }
\end{tikzpicture}%
\end{NiceTabular}%
\end{minipage}%
\label{tbl:parameter_prediction}%
}\\
\subfloat[Root mean square error in predicted $\alpha_{\text{CIL}}$ and $\alpha_{\text{CF}}$ from neural network trained on $\featFfull{}$.]{%
\begin{minipage}{\textwidth}
\centering
\includegraphics[]{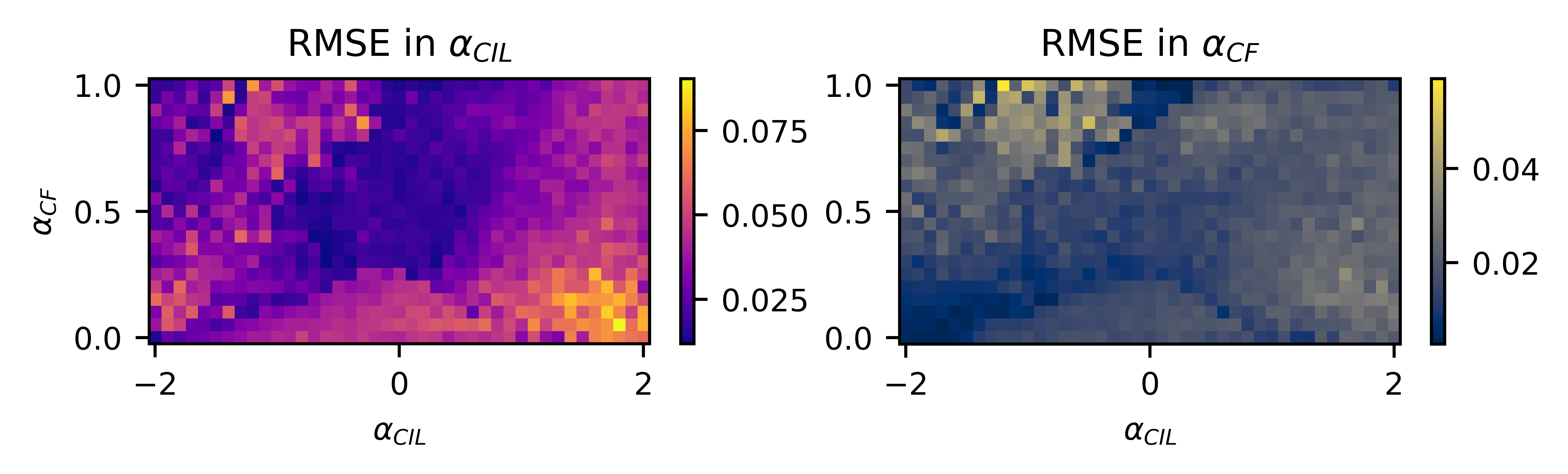}%
\end{minipage}%
\label{fig:parameter_prediction_rmse}%
}%
\caption{Recovering interaction parameters $\alpha_{\text{CIL}}$ and $\alpha_{\text{CF}}$ from extracted motif features using a fully connected neural network model.
The model consists of three hidden layers with 2048, 2048 and 8 neurons and uses the $\operatorname{ReLU}$ activation function used between these hidden layers.
Standard deviation in error are $0.0373$ and $0.0201$ for $\alpha_{\text{CIL}}$ and $\alpha_{\text{CF}}$ respectively.}%
\label{fig:parameter_prediction}%
\end{figure*}

Here, 861 movies of cell trajectories were generated at unique pairs of $\alpha_{\text{CIL}}$ and $\alpha_{\text{CF}}$.
From each movie, motif features were extracted from 960 distinct snapshots after their DSO patterns have appeared.
We coarse-grained the feature vectors $\featEfull{}$ over multiple particles at the snapshot into $861 \times 960 = 826,560$ sets of 42-dimensional feature vectors $\featFfull{}$.
These 42-dimensional feature vectors $\featFfull{}$ are the input to the neural network trained to predict the corresponding $\alpha_{\text{CIL}}$ and $\alpha_{\text{CF}}$ parameters (see \cref{fig:dimensionful_param_loss}).
Because these input vectors were sufficiently telling, we only needed to train a relatively simple multi-layer perceptron, which consists of three hidden layers (with 2048, 2048 and 8 neurons in each respective layer) that uses the $\operatorname{ReLU}$ activation function.

\Cref{fig:parameter_prediction_rmse} showed that this neural network recovered the interaction parameters with reasonable accuracies.
For validation, we extracted motif feature-vectors from 240 additional snapshots from each movie that were only used to test and not train the neural network.
The root mean squared error in prediction of these previously unseen test samples is generally small, at $0.0373$ for $\alpha_{\text{CIL}}$ and $0.0201$ for $\alpha_{\text{CF}}$.
We note that the relative error in $\alpha_{\text{CIL}}$ is smaller than the relative error in $\alpha_{\text{CF}}$.

Abstractly, this neural network is made to map the 42-dimensional input features into an 8-dimensional manifold that could be used to predict the $\alpha_{\text{CIL}}$ and $\alpha_{\text{CF}}$ parameters.
This manifold, which is imbued with information from both inputs and outputs, are often interrogated for insights. 

\begin{figure*}[!t]
\centering
\includegraphics[]{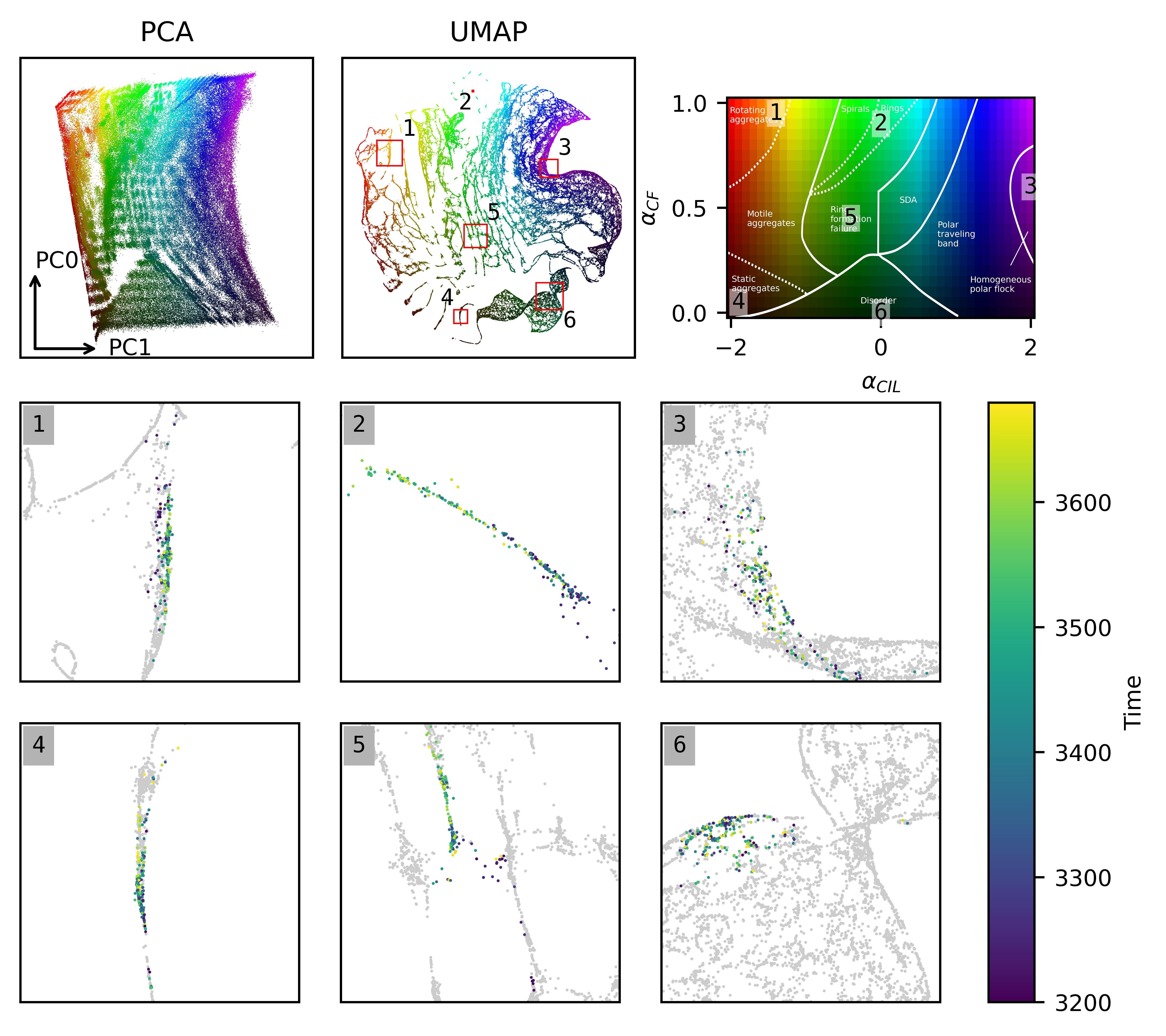}%
\caption{Embedding space learned by the neural network in \cref{fig:parameter_prediction}.
Each point represents $\featBfull{}$, each of which corresponds to a single time point of a movie.
(TOP) The network transforms $\featBfull{}$ into the eight-dimensional output of the network's last hidden layer, which we embed into two dimension either through PCA or UMAP.
These embeddings are colourised by the corresponding $\alpha_{\text{CIL}}$ and $\alpha_{\text{CF}}$ (legend on the top right).
We drew red boxes in the UMAP embedding that encompasses the points from six movies (enumerated in the legend), which were magnified in the panels below.
(BOTTOM) These magnified panels were colourised by their respective times in the movie, with points from other movies in grey.
Panels 1, 2, 4 and 6 show movie instances with features that are locally constrained and vary as low-dimensional `strands'.
The embedded points from these movies also evolve continuously along these `strands'.
At $\alpha_{\text{CIL}}<0$, features tend to be locally constrained vary as low-dimensional `strands', and embedded points from each movie often evolve continuously along a single `strand'.
}%
\label{fig:dimensionful_embedding_phases}%
\end{figure*}

In \cref{fig:dimensionful_embedding_phases}, we embed this manifold into two dimensions using Principal Component Analysis (PCA).
The two principal components of this embedding reproduces the two dimensional $\alpha_{\text{CIL}}$ and $\alpha_{\text{CF}}$ parameter space on the top right remarkably well.
Further, most of the variance in this embedding lies in the $\alpha_{\text{CF}}$ direction.
The embedding also shows some degree of separation between points from the disordered pattern and the other patterns.

To better resolve the structures in this manifold, we also show its UMAP embedding in \cref{fig:dimensionful_embedding_phases}.
While in the PCA embedding, the disordered DSO patterns appear distinct from the other ones, this distinction becomes abundantly clear in the UMAP embedding.
The additional structure in the UMAP embedding also indicates that frames in the $\alpha_{\text{CIL}} > 0$ regions are more connected than those in the $\alpha_{\text{CIL}} < 0$ regions.
Movies corresponding to the more connected or compressed areas of the UMAP embedding (e.g., disordered, polar flock, travelling band, etc.) tends to have larger root mean squared error (\Cref{fig:parameter_prediction_rmse}).
However, we caution against over-interpreting these embeddings which could suffer topological instabilities from having insufficient data samples.

Even though this neural network was explicitly trained to predict the interaction parameters from the $\featFfull{}$ inputs alone, it is also able to distinguish more nuances in these inputs compared to the transformation and classification performed in \cref{phase_space}.
This increased distinguishability likely comes from the trained network's increased sensitivity for certain non-linear combinations of the features in the input.
The resultant fine structure in the embedding of the network's transformation of these inputs arise from emphasizing these non-linear feature combinations.

To better appreciate this embedding's fine structure, we examine the features vectors of frames taken from six different movies (each with their own unique pairs of $\alpha_{\text{CIL}}$ and $\alpha_{\text{CF}}$ interaction parameters.
Consider the relatively isolated embedding from the frames of a particular movie of rotating aggregates (panel 1).
The neural network essentially learned which non-linear combination of these frames' feature vectors $\featFfull{}$ to `pay attention to' in order to discern the interaction parameters.
The last layer of the neural network does this by mapping the input feature vector to an isolated region in the manifold.
Similar mappings also apply to static aggregates (panel 4) and ring formation failure (panel 5).
Additionally, the frames from spirals (panel 2) appear to also be time ordered, evident from the colouration of their corresponding points according to time in the movies.

In contrast, feature vectors from the disordered phase (panel 6 of \cref{fig:dimensionful_embedding_phases}) tend to be more scattered.
Evidently, the embedding of these frames also tends to overlap with those in other disordered movies with different interaction parameters (i.e., grey points in the figure).
Since the neural network maps points on this embedding to specific pairs of $\alpha_{\text{CIL}}$ and $\alpha_{\text{CF}}$ parameters, this overlap shows how the neural network erroneously maps frames from different movies to the same parameter pairs.
The same applies for homogeneous polar flocks in panel 3.

When extracting motif features from each movie, feature categories dependent on the length and time scales have been included.
Comparing the features obtained from different systems would require these scaling to be matched, which can be difficult to do in practice.

To address this issue, we retrained a neural network in \cref{fig:dimensionless_param_loss} to predict each frame's $\alpha_{\text{CIL}}$ and $\alpha_{\text{CF}}$ using only the dimensionless features $\featHfull{}$ in \cref{tbl:dimless_parameter_prediction} as input.
While the architecture of this second network is identical to the previous case except for the input layer, \cref{fig:dimless_parameter_prediction_rmse} shows this network makes less accurate predictions, with root mean squared error of $\alpha_{\text{CIL}}$ for $0.147$ and $\alpha_{\text{CF}}$ for $0.0434$.
This error is most evident in the prediction of $\alpha_{\text{CF}}$ for rings and motile aggregates.

%dimless_parameter_prediction

%Compared to embedding using all motif features, points from each movie are more scattered and mixed together with other phases.
\Cref{fig:dimensionless_embedding_phases} shows the embedding the manifold learned by the neural network trained on dimensionless features.
Whereas this PCA embedding still shows hints of a two-dimensional manifold, there is now a pronounced separation between the disordered DSO pattern and the other patterns.

%dimensionless_embedding_phases

Unlike the embedding from \cref{fig:dimensionful_embedding_phases}, most movies in the manifold form connected patches, and several (e.g., frames from movies shown in panel 1, 2, 4 and 5) have frames with embeddings that are widely scattered.
This connectedness is consistent with the large RMSE for this network compared to the previous one, and shows the importance of the dimension-full features that were present in the former network (mean neighbour distance, motif speed, velocity dot product, divergence and curl) for predicting the interaction parameters.

%% file: sections/phase_classification_figure.tex
\setlength{\tableJoinSep}{5pt}

\begin{figure*}[t]
\centering
\subfloat[Features $\featAtbl{}$ coarse-grained over all cells $n$ at each time $t$ in six different ways to get $\featBtbl{}$, and then over all time into 72-dimensional feature vectors $\featCtbl{}$. UMAP embedding of the $\featCtbl{}$ vectors are shown as points and clustered on the right.]{%
\begin{minipage}{0.455\textwidth}
\classiffontsize{}%
\centering
\begin{NiceTabular}[t]{|c|@{}p{\tableJoinSep}@{}|c|@{}p{\tableJoinSep}@{}c}
\hhline{-~-~-}
$\featAtbl{} \in \mathbb{R}^{6}$ && $\featBtbl{} \in \mathbb{R}^{36}$ & \multicolumn{1}{@{}c@{}|}{} & \multicolumn{1}{c|}{$\featCtbl{} \in \mathbb{R}^{72}$}\\\hhline{=~=~|=|}
Neighbour Count      && $\underset{n}{\operatorname{min}}{\, \featAtbl{}}$  & \multicolumn{1}{@{}c@{}|}{} & \multicolumn{1}{c|}{$\underset{t}{\operatorname{mean}}{\, \featBtbl{}}$}\\\hhline{-~-~|-|}
Mean Distance        && $\underset{n}{\operatorname{max}}{\, \featAtbl{}}$  & \multicolumn{1}{@{}c@{}|}{} & \multicolumn{1}{c|}{$\underset{t}{\operatorname{std}}{\, \featBtbl{}}$}\\\hhline{-~-~-}
Centre Speed         && $\underset{n}{\operatorname{mean}}{\, \featAtbl{}}$ &\\\hhline{-~-~~}
Velocity Dot Product && $\underset{n}{\operatorname{var}}{\, \featAtbl{}}$  &\\\hhline{-~-~~}
Divergence           && $\underset{n}{\operatorname{skew}}{\, \featAtbl{}}$ &\\\hhline{-~-~~}
Curl                 && $\underset{n}{\operatorname{kurt}}{\, \featAtbl{}}$ &\\\hhline{-~-~~}
\CodeAfter
\begin{tikzpicture}
    \foreach \x in {2.5,...,7.5}{
        \draw (\x-|2) -- (2.5-|3);
        \draw (\x-|4) -- (2.5-|5);
        \foreach \y in {3.5,...,7.5}{
            \draw[color=black!25] (\x-|2) -- (\y-|3);
        }
        \draw[color=black!25] (\x-|4) -- (3.5-|5);
    }
\end{tikzpicture}%
\end{NiceTabular}%
\end{minipage}%
\begin{minipage}{0.545\textwidth}
\includegraphics[valign=t]{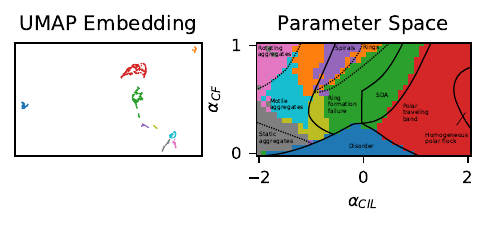}%
\end{minipage}%
%\label{tbl:movie_phase_cluster}%
\label{fig:movie_phase_cluster}%
}\\
\subfloat[Coarse-graining $\featAtbl{}$ over all $n$ cells at each time into 36-dimensional feature vectors $\featBtbl{}$.]{%
\begin{minipage}{0.455\textwidth}
\classiffontsize{}%
\centering
\begin{NiceTabular}[t]{|c|@{}p{\tableJoinSep}@{}|>$c<$|}
\hhline{-~-}
$\featAtbl{} \in \mathbb{R}^{6}$ && \featBtbl{} \in \mathbb{R}^{36}\\\hhline{=|~|=|}
Neighbour Count      && \underset{n}{\operatorname{min}}{\, \featAtbl{}}\\\hhline{-~-}
Mean Distance        && \underset{n}{\operatorname{max}}{\, \featAtbl{}}\\\hhline{-~-}
Centre Speed         && \underset{n}{\operatorname{mean}}{\, \featAtbl{}}\\\hhline{-~-}
Velocity Dot Product && \underset{n}{\operatorname{var}}{\, \featAtbl{}}\\\hhline{-~-}
Divergence           && \underset{n}{\operatorname{skew}}{\, \featAtbl{}}\\\hhline{-~-}
Curl                 && \underset{n}{\operatorname{kurt}}{\, \featAtbl{}}\\\hhline{-~-}
\CodeAfter
\begin{tikzpicture}
    \foreach \x in {2.5,...,7.5}{
        \draw (\x-|2) -- (2.5-|3);
        \foreach \y in {3.5,...,7.5}{
            \draw[color=black!25] (\x-|2) -- (\y-|3);
        }
    }
\end{tikzpicture}%
\end{NiceTabular}%
\end{minipage}%
\begin{minipage}{0.545\textwidth}
\includegraphics[valign=t]{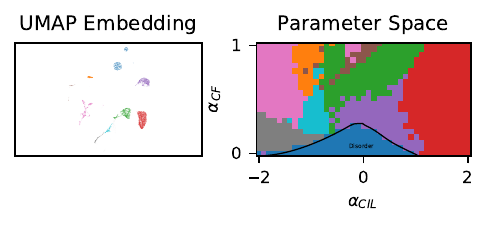}%
\end{minipage}%
%\label{tbl:frame_phase_cluster}%
\label{fig:frame_phase_cluster}%
}\\
\subfloat[Features $\featAtbl{}$ coarse-grained over time into vectors of 30 features $\featDtbl{}$ to uncover `phases'.]{%
\begin{minipage}{0.455\textwidth}
\classiffontsize{}%
\centering
\begin{NiceTabular}[t]{c@{}p{\tableJoinSep}@{}|>$c<$|}
\hhline{-~-}
\multicolumn{1}{|c|}{$\featAtbl{} \in \mathbb{R}^{5}$} && \featDtbl{} \in \mathbb{R}^{30}\\\hhline{|=|~=}
\multicolumn{1}{|c|}{Mean Distance}        && \underset{t}{\operatorname{min}}{\, \featAtbl{}}\\\hhline{-~-}
\multicolumn{1}{|c|}{Centre Speed}         && \underset{t}{\operatorname{max}}{\, \featAtbl{}}\\\hhline{-~-}
\multicolumn{1}{|c|}{Velocity Dot Product} && \underset{t}{\operatorname{mean}}{\, \featAtbl{}}\\\hhline{-~-}
\multicolumn{1}{|c|}{Divergence}           && \underset{t}{\operatorname{var}}{\, \featAtbl{}}\\\hhline{-~-}
\multicolumn{1}{|c|}{Curl}                 && \underset{t}{\operatorname{skew}}{\, \featAtbl{}}\\\hhline{-~-}
                                           && \underset{t}{\operatorname{kurt}}{\, \featAtbl{}}\\\hhline{~~-}
\CodeAfter
\begin{tikzpicture}
    \foreach \x in {2.5,...,6.5}{
        \draw (\x-|2) -- (2.5-|3);
        \foreach \y in {3.5,...,7.5}{
            \draw[color=black!25] (\x-|2) -- (\y-|3);
        }
    }
\end{tikzpicture}%
\end{NiceTabular}%
\end{minipage}%
\begin{minipage}{0.545\textwidth}
\includegraphics[valign=t]{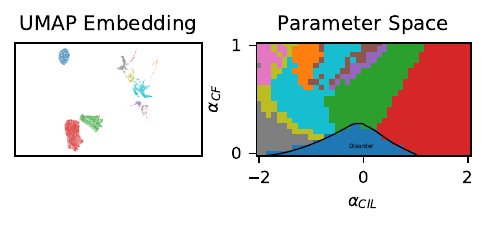}%
\end{minipage}%
%\label{tbl:group_phase_cluster}%
\label{fig:group_phase_cluster}%
}%
\caption{`Phase' spaces uncovered from coarse-graining structural and dynamic motif features show differences between one another. The tables shows the three different ways motif features of the $n^{\mathrm{th}}$ cell are summarised to obtain three sets of coarse-grained features.
From each set of coarse-grained features, a `phase' spaces is uncovered by embedding the feature vectors as two-dimensional points using UMAP and clustering the points using density-based clustering (HDBSCAN). The clustered points are coloured by their most likely cluster label in the embedding space; points without valid labels are light grey.}%
\label{fig:phase_cluster}
\end{figure*}

%% file: sections/conclusion.tex
\section{Conclusion}
\label{conclusion}
In complex systems with many interacting objects, the objects can spontaneously form groups that exhibit prevalent and persistent patterns, which we refer to as motifs.
These motifs can possess quantitative features that encapsulate how a group of objects interact with one another, enabling us to describe their interactions efficiently.

When patterns emerge from interactions between motifs, higher-level motifs can be built.
Iterating this motif-building process results in a hierarchy of motifs and motif features that characterise a system over multiple length scales.

Using motifs and their features, we analysed the collective structures and dynamics that emerge in simulations of Dynamically Self-Organising (DSO) cells.

Within large compact cell collectives (e.g., motile and rotating aggregates) formed when cells `attract' (i.e., $\alpha_{\text{CIL}} < 0$), we observed strain and defects in the cell packing.
Attracted inwards of the collective, cells packed into increasingly strained and compressed 6NN motifs towards the centre of the cell collective.
Meanwhile, the packing defects appeared as dynamically fluctuating chains of alternating 5NNs and 7NNs.
Such observations show how motifs can reveal structural details of object collectives within complex systems.

Similarly, we uncovered two dynamic properties of cell collectives: First, that the centre of aggregates (i.e., static, motile and rotating) contains actively moving cells; Second, that spirals evolved in a semi-periodic manner between two modes.
In aggregates, these active cell movements were correlated with a longer motif lifetime in motile and rotating aggregates, which have stable interior cell circulation.
We also found that motifs tend to be long-lived when they are situated in the slow-moving regions of the aggregate.
In spirals, their temporal semi-periodic alternation between an elongated and a rotund form was reflected as motif features, the mean neighbour distance in this case, whose distribution also alternates with time.
These insights into the collective movements and evolution of cell collectives highlight that motif features can uncover emergent dynamics in groups of objects.

Using the richness of the motif features, we employed them in machine learning to classify or recover the hidden interaction parameters ($\alpha_{\text{CIL}}$ and $\alpha_{\text{CF}}$) from simulated cell trajectories.

To classify the interaction parameter phase space, we performed unsupervised clustering on suitably coarse-grained motif features from all cell groups in a movie.
The resultant clustering appeared similar to the DSO phase diagram identified in Ref. \cite{hiraiwaDynamicSelfOrganizationIdealized2020}.
We further tested the robustness of the classes by imposing limitations on the extraction of motif features in one of two ways: either the features are extracted from one point in time, or the features are extracted from motifs centred on one particular cell as it evolves over time.
All three obtained clusterings showed similar `phase boundaries' in certain DSO pattern regions, particularly in the weakly interacting disordered region.
This result suggests that motifs can give robust features for classifying interactions between objects in experiments, even if the experimental data have limited time samples or track only a few objects.

Similarly, we showed that motif features extracted at a single time point of a system can recover the system's hidden $\alpha_{\text{CIL}}$ and $\alpha_{\text{CF}}$ interaction parameters.
By training a simple fully connected neural network model on the summary statistics of these motif features, we obtained predictions with a root mean squared error of 0.0373 for $\alpha_{\text{CIL}}$ and 0.0201 for $\alpha_{\text{CF}}$.
Notably, the model achieves an error approximately half of the interval between adjacent $\alpha_{\text{CIL}}$ and $\alpha_{\text{CF}}$ values used for training the model.
From the latent embedding learnt by the neural network, we found that the motif features tend to change continuously over time as the system evolves.
Such predictions demonstrate how motif features can effectively address the inverse problem of aligning observations in experimental systems with parameters of a theoretical model.

Finally, we identified a higher-level motif (dipolar vortices) in cell aggregates whose features predicted the aggregate's movement and rotation.
These dipolar vortices motifs were formed by hierarchically coarse-graining neighbouring cell group motifs that rotated in the same manner.
The positions of these dipolar vortices gave robust predictions on the aggregates' future velocity, while their relative rotation strength correlated with the aggregates' angular velocity.

Although motif features are not real thermodynamic variables and should not be over-relied on as objective metrics, we believe they may offer a theoretical framework for studying complex systems.
Further research applying this framework to other theoretical and experimental systems would help validate its generalisability.
Additionally, the potential of using motifs to study generic systems can be enhanced by discovering new ways to coarse-grain low-level motifs to higher-level motifs with less human supervision.

%% file: arxiv/SI.tex
\subsection{Mathematical Model for Dynamic Self-Organisation of Migrating Cells through Intracellular Contact Communication}
\label{SI:model}
Here we briefly review the mathematical model \cite{hiraiwaDynamicSelfOrganizationIdealized2020} employed in this study as a test model.

Each migrating cell is approximated as a self-propelled particle moving in two dimensions with contact interactions in the form of volume exclusion and the effects of cell-cell communication.
The position $\boldsymbol{x}_j$ and the vector representing intrinsic polarity $\boldsymbol{q}_j$ of the $j$-th cell evolves in time obeying the following equations of motion:
\begin{equation}
\boldsymbol{v}_j = \odv{\boldsymbol{x}_j}{t} = v_0 \boldsymbol{q}_j + \boldsymbol{J}^\nu_j
\end{equation}
\begin{equation}
\odv{\theta_j}{t} = \boldsymbol{J}^q_j \cdot \boldsymbol{q}_{j,\perp} + \xi_j
\end{equation}

The polarity of each cell has a fixed magnitude and is given by $\boldsymbol{q}_j=(\cos \theta_j,\sin \theta_j)$, with its perpendicular direction by $\boldsymbol{q}_{j,\perp}=(-\sin \theta_j,\cos \theta_j)$.
The coefficient $v_0$ represents the migration speed of a cell in the absence of volume exclusion.

Cells are modelled as soft disks of radius $r$, and their volume exclusion is described by the term
\begin{equation}
\boldsymbol{J}^\nu_j = -\beta \sum_{j' \in N_j} \left(r\left|\boldsymbol{\Delta x}_{j',j}\right|^{-1} - 1\right) \widehat{\boldsymbol{\Delta x}_{j',j}}
\end{equation}
where $\boldsymbol{\Delta x}_{j',j} = \boldsymbol{x}_{j'} - \boldsymbol{x}_j$ and $\widehat{\boldsymbol{\Delta x}_{j',j}} = {\boldsymbol{\Delta x}_{j',j}} / {\left|\boldsymbol{\Delta x}_{j',j}\right|}$.
The summation, $\sum_{j' \in N_j} \square$, runs over all neighbouring cells $N_j$ that satisfies $\left|\boldsymbol{\Delta x}_{j',j}\right| < r$.
The coefficient $\beta$ represents the interaction strength of the soft-disk volume exclusion.

The intercellular communication term $\boldsymbol{J}^q_j$ comprises two parts: contact following (CF) and contact inhibition/attraction of locomotion (CIL/CAL),
\begin{equation}
\boldsymbol{J}^q_j = \boldsymbol{J}^\text{CF}_j + \boldsymbol{J}^\text{CIL}_j \ .
\end{equation}
The CF term is given by
\begin{equation}
\boldsymbol{J}^\text{CF}_j = \alpha_{\text{CF}} \sum_{j' \in N_j} \frac{1 + \widehat{\boldsymbol{\Delta x}_{j',j}} \cdot \boldsymbol{q}_{j'}}{2} \widehat{\boldsymbol{\Delta x}_{j',j}}
\end{equation}
and the CIL/CAL term is given by
\begin{equation}
\boldsymbol{J}^\text{CIL}_j = -\alpha_{\text{CIL}} \sum_{j' \in N_j} \left(r\left|\boldsymbol{\Delta x}_{j',j}\right|^{-1} - 1\right) \widehat{\boldsymbol{\Delta x}_{j',j}}
\end{equation}
The coefficient $\alpha_{\text{CF}}$ represents the interaction strength for CF and is non-negative.
The coefficient $\alpha_{\text{CIL}}$ represents the CIL/CAL strength, where $\alpha_{\text{CIL}} > 0$ corresponds to CIL while $\alpha_{\text{CIL}} < 0$ corresponds to CAL.

The last term, $\xi_j$, represents Gaussian white noise satisfying $\left<\xi_j\right> = 0$ and $\left<\xi_j(t)\xi_{j'}(t')\right> = 2D\delta_{jj'}\delta(t-t')$, where $D$ is the noise intensity.

\subsection{Supplementary Figures}
%\begin{figure}[H]
%	\includegraphics[]{motif_lifetime}%
%	\caption{Motif stability increases with higher contact following strength $\alpha_{\text{CF}}$, organising cells into patterns that have longer motif lifetime.}%
%	\label{fig:motif_lifetime}%
%\end{figure}

\begin{figure}[H]
\centering
    \includegraphics[]{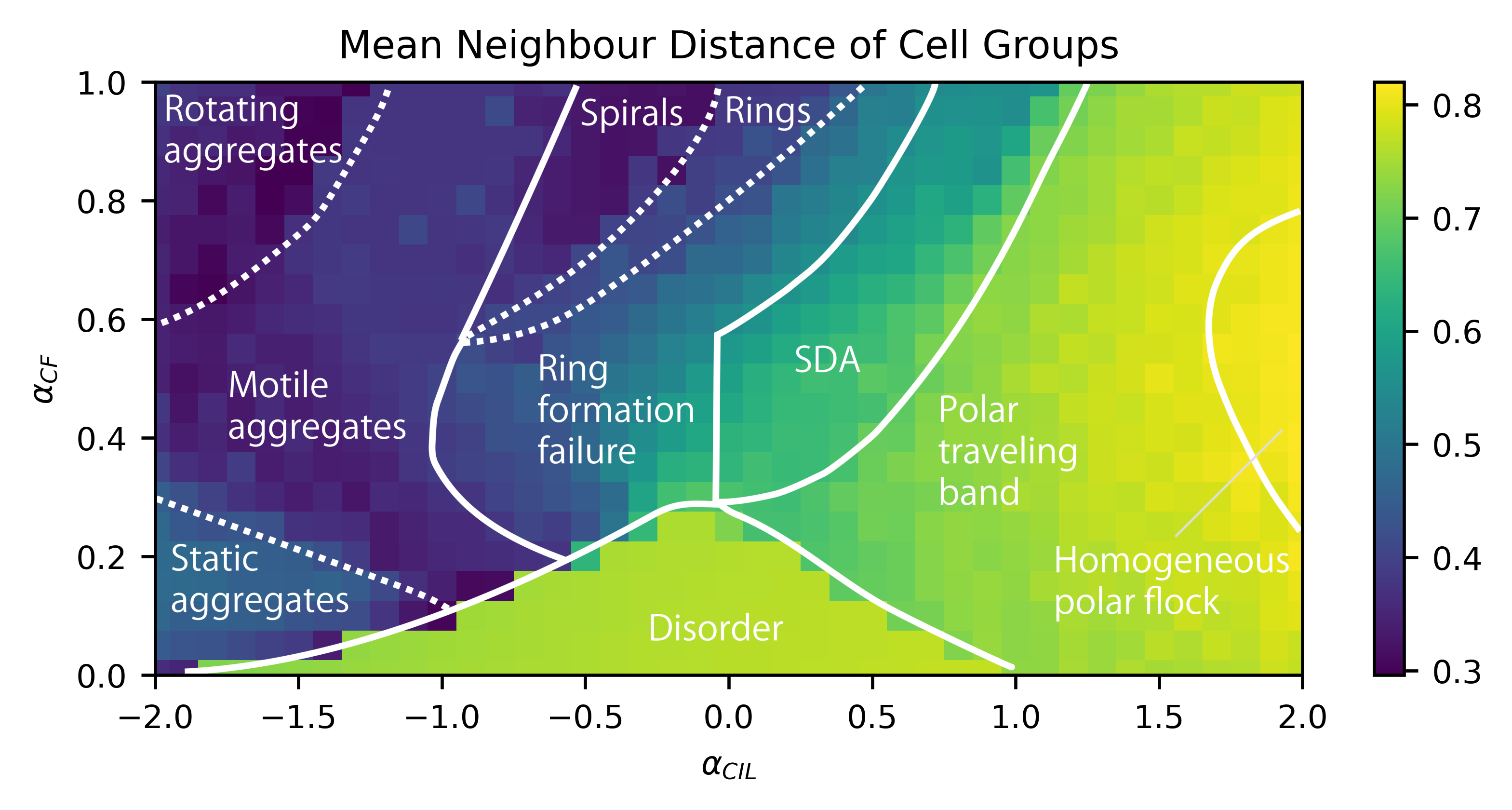}%
    \caption{Structural feature of cell groups showing `phases' of collective behaviour in simulated dynamically self-organising (DSO) cells. The parameter space is sectioned into phases based on the movement patterns of cell collectives \cite{hiraiwaDynamicSelfOrganizationIdealized2020}.}%
    \label{fig:dso_phase}%
\end{figure}

\begin{figure}[H]
\centering
    \includegraphics[]{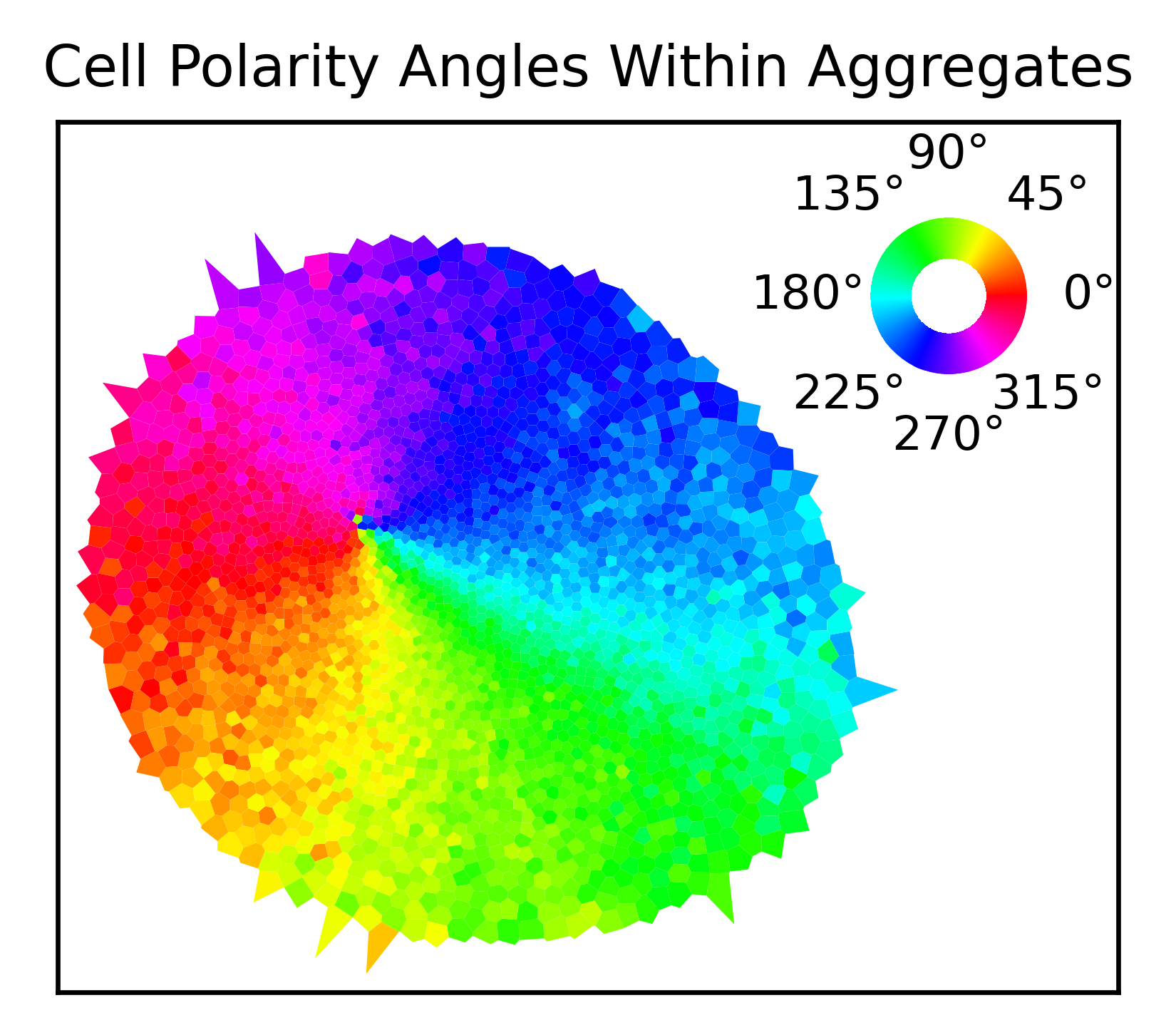}%
    \caption{Dense, motile aggregates have cells with inwards pointing polarity. Cells are coloured by the direction of their polarity according to the colour wheel.}%
    \label{fig:aggregate_polarity}%
\end{figure}

\begin{figure}[H]
\centering
    \includegraphics[]{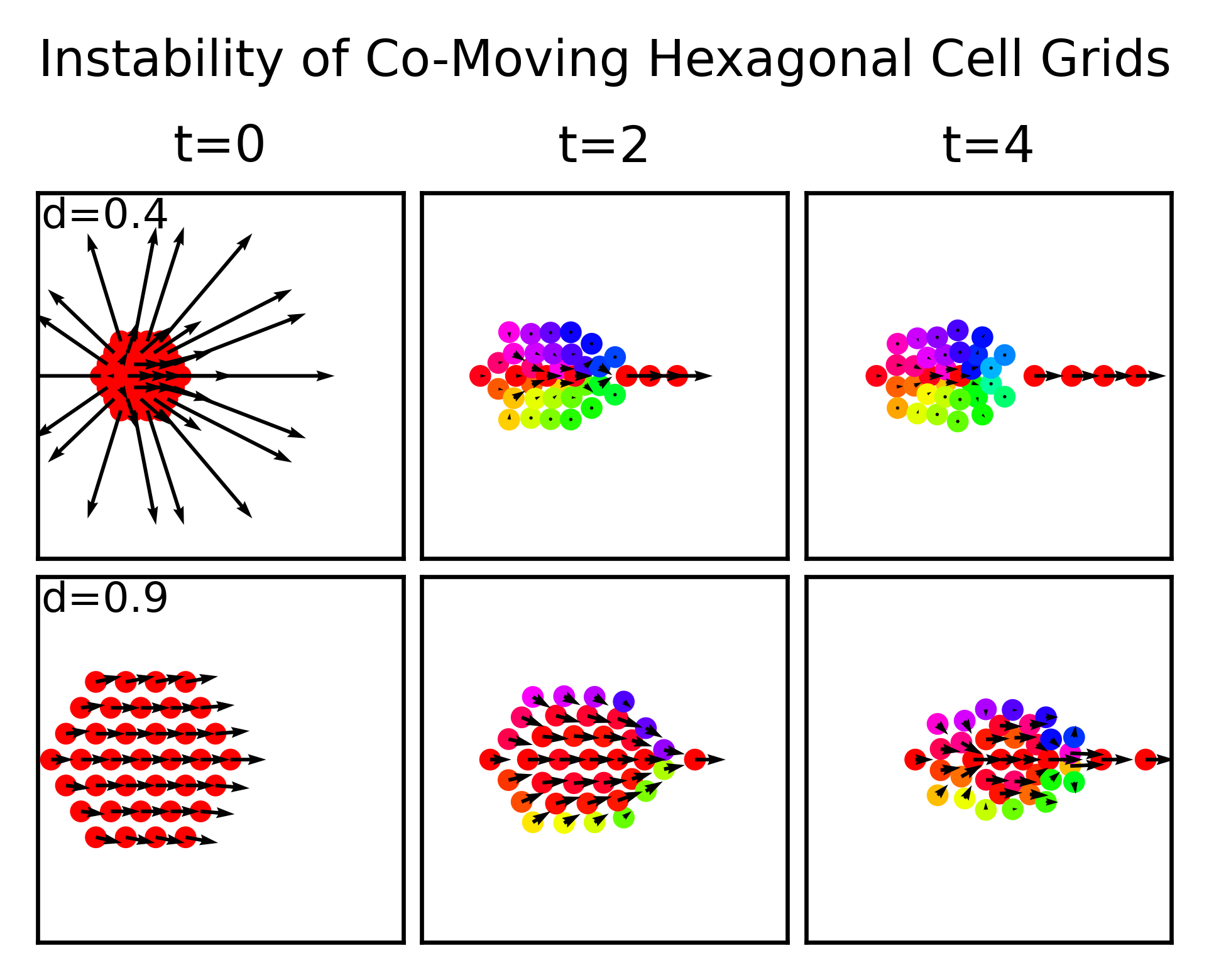}%
    \caption{Isolated hexagonal cell patches with co-aligned polarity from motile aggregates are unstable.
Cells are coloured according to their polarity directions; arrows display each cell's velocity.
The polarity of the cells will turn towards the patch's centre of mass, forming a new aggregate with inter-cell distances dependent on the aggregate's size.}%
    \label{fig:unstable_comoving_hex}%
\end{figure}

\begin{figure}[H]
\centering
    \includegraphics[width=\textwidth]{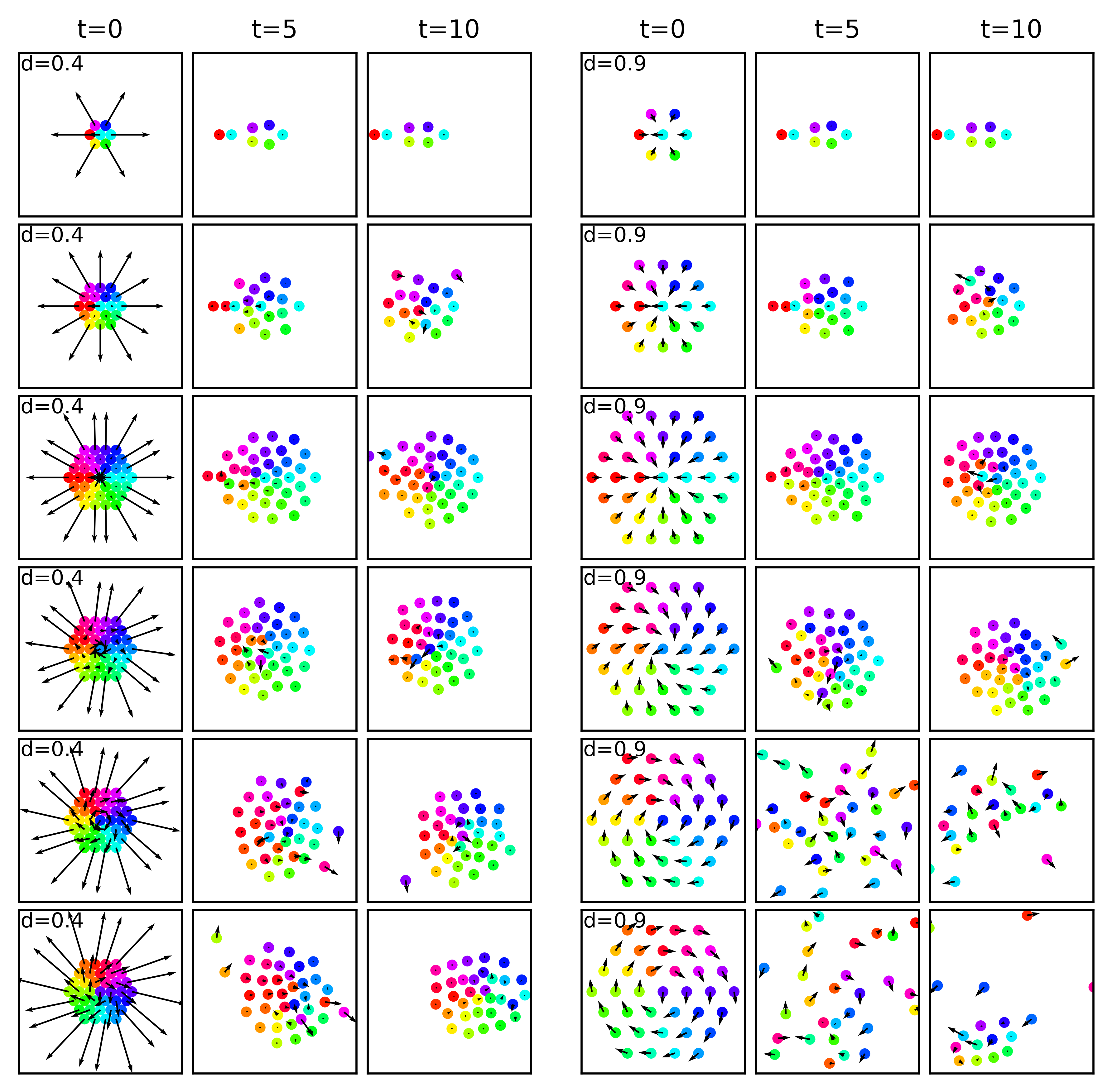}%
    \caption{Cell from the motile aggregate Dynamic Self-Organised (DSO) pattern initialised in a hexagonal grid with inwards pointing polarity and different initial separation $d$. Cells are coloured by their internal polarity angle, with arrows showing their current velocity. Offsets in initial polarity are corrected quicker when cell separation is small, and aggregates appear more stable when more cells are present.}%
    \label{fig:PersistentAggregate}%
\end{figure}

\begin{figure}[H]
\centering
	\includegraphics[]{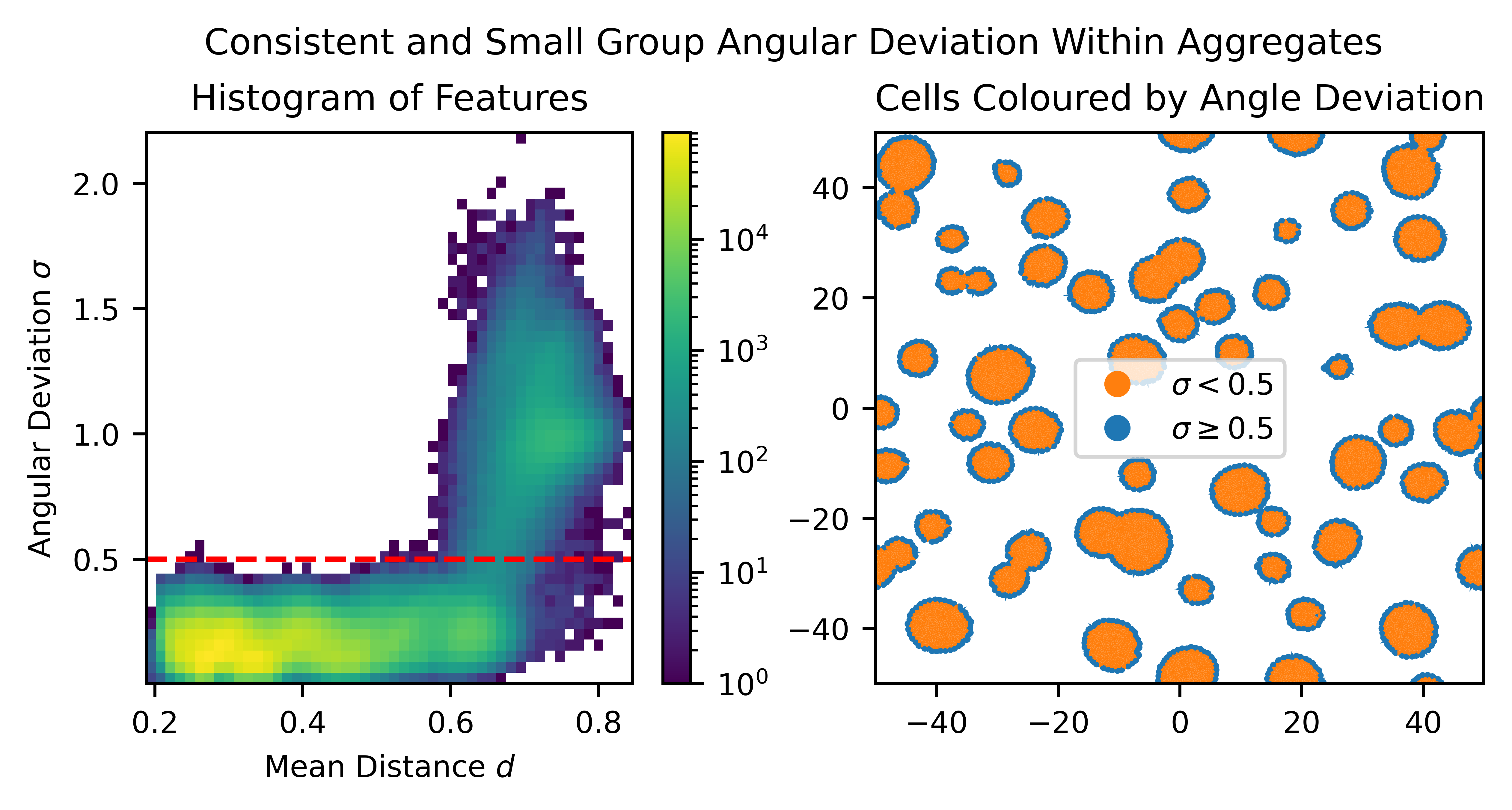}%
	\caption{Formation of different types of cell groups at interior and surface of aggregates.
 (Left) Cell groups within aggregates are compressible while maintaining a small range of angular deviation.
 (Right) Low angular deviation corresponds with lying within aggregates.}%
	\label{fig:dense_sparse_cutoff}%
\end{figure}

\begin{figure}[H]
\centering
    \includegraphics[]{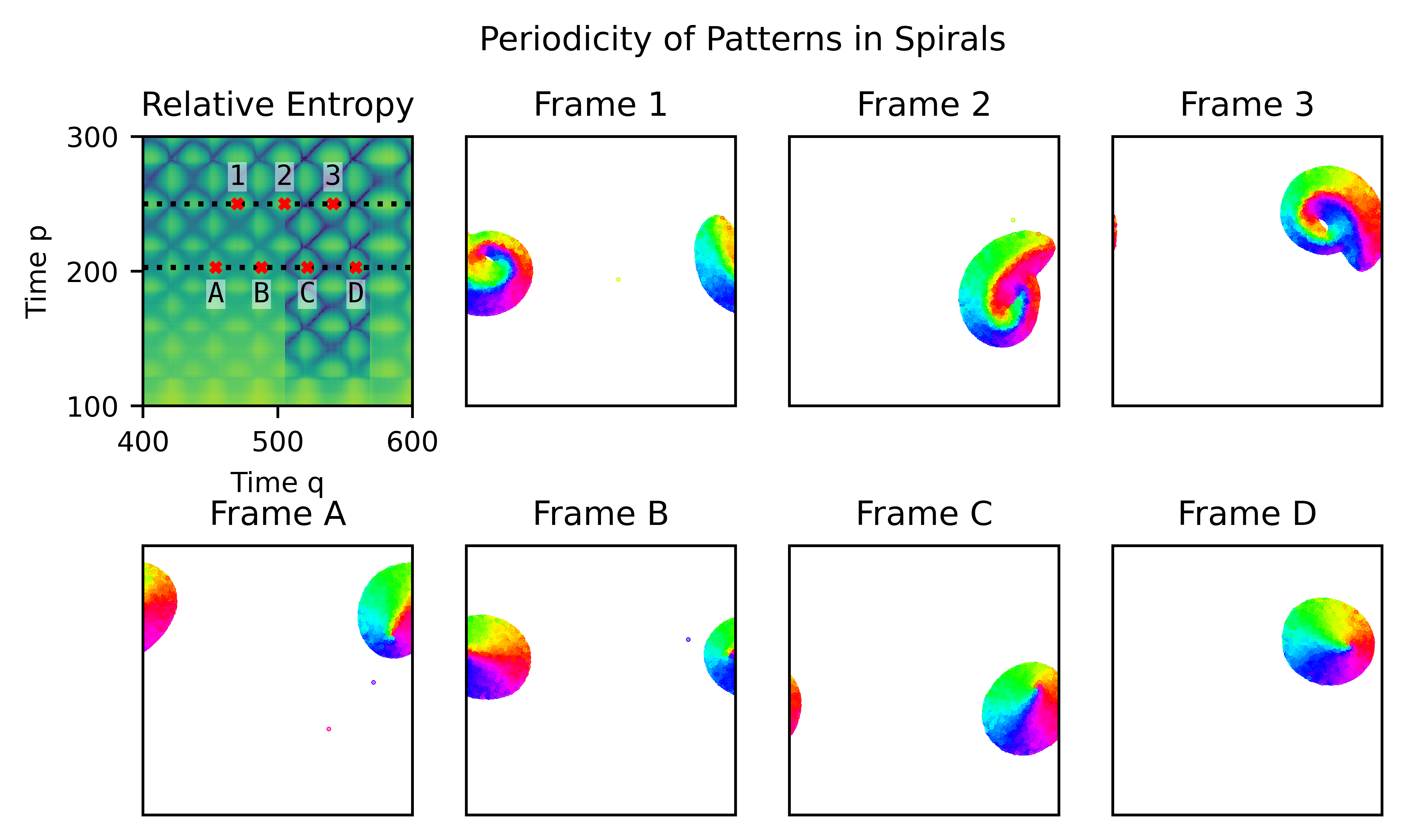}%
    \caption{Periodicity of relative entropy corresponds to different cell organisation modes in spirals. The relative entropy compares the mean neighbour distance distribution of cell groups between different time points.}%
    \label{fig:spiral_periodic}%
\end{figure}

\begin{figure}[H]
\centering
    \includegraphics[]{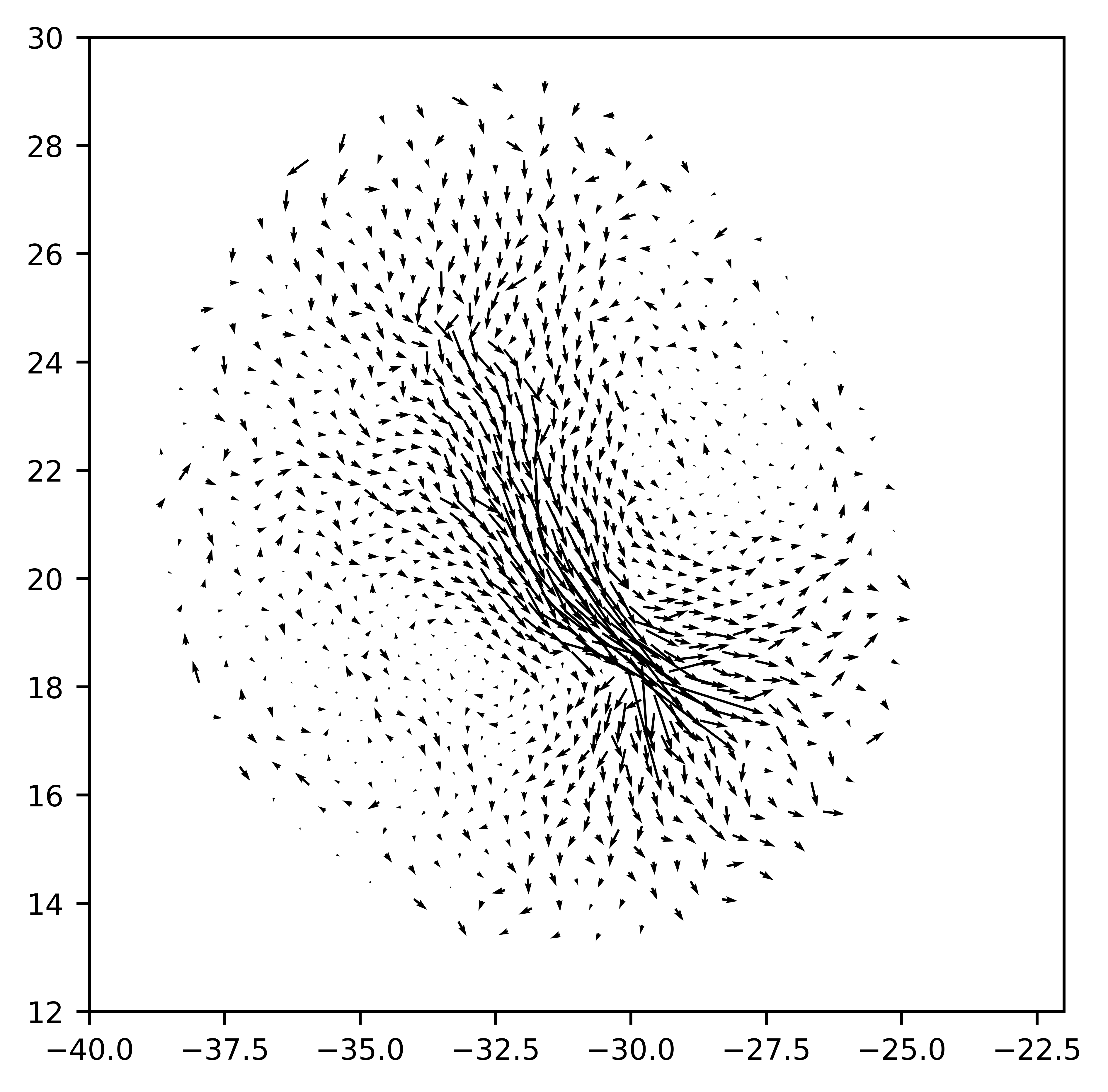}%
	\caption{Velocity of each cell circulating within the aggregate highlighted in \cref{fig:movement_prediction}. The cells move quickly along the central stream before slowing down as it exits the stream, creating a net movement in the aggregate.}%
	\label{fig:aggregate_cell_velocity}%
\end{figure}

\begin{figure}[H]
\centering
    \includegraphics[]{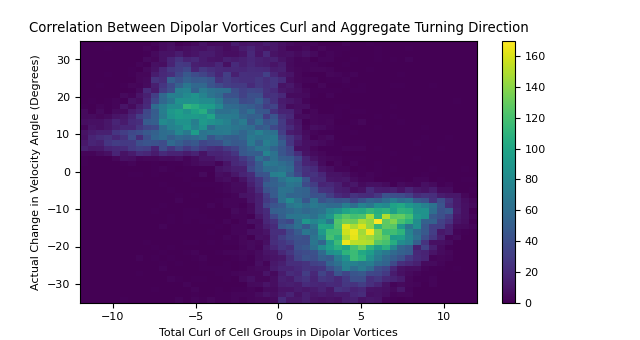}%
    \caption{Larger total curl magnitude in the positive vortex (cell-groups labelled red in \cref{fig:movement_prediction}) correlates with aggregates turning clockwise (i.e., a decrease in velocity angle).}%
    \label{fig:aggregate_turning}%
\end{figure}

\begin{figure}[H]
\centering
\includegraphics[]{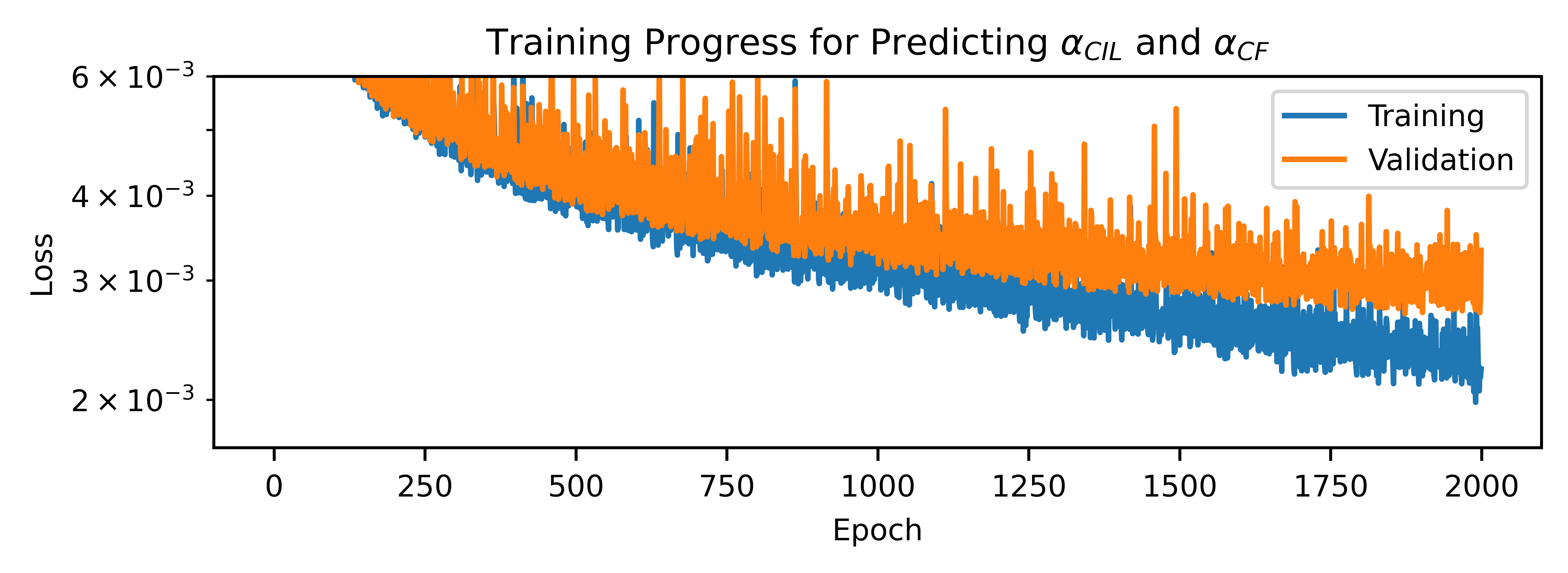}%
\caption{Mean squared error (loss) of a neural network decreasing while it trains to make predictions of $\alpha_{\text{CIL}}$ and $\alpha_{\text{CF}}$ from motif features.
From movies with unique pairs of $\alpha_{\text{CIL}}$ and $\alpha_{\text{CF}}$ parameter, motif features of cells in each movie frame were coarse-grained into 42-dimensional vectors and used as features for predicting their corresponding $\alpha_{\text{CIL}}$ and $\alpha_{\text{CF}}$.
The neural network has fully connected hidden layers of size 2048, 2048 and 8, and uses the ReLU activation function.
Testing the network on previously unseen feature samples gives a test loss of 0.00269
Note that the loss here is computed against the $\alpha_{\text{CIL}}$ and $\alpha_{\text{CF}}$ parameters after the parameters were scaled to have a standard deviation of 1.}%
\label{fig:dimensionful_param_loss}%
\end{figure}

\begin{figure}[H]
\centering
\includegraphics[]{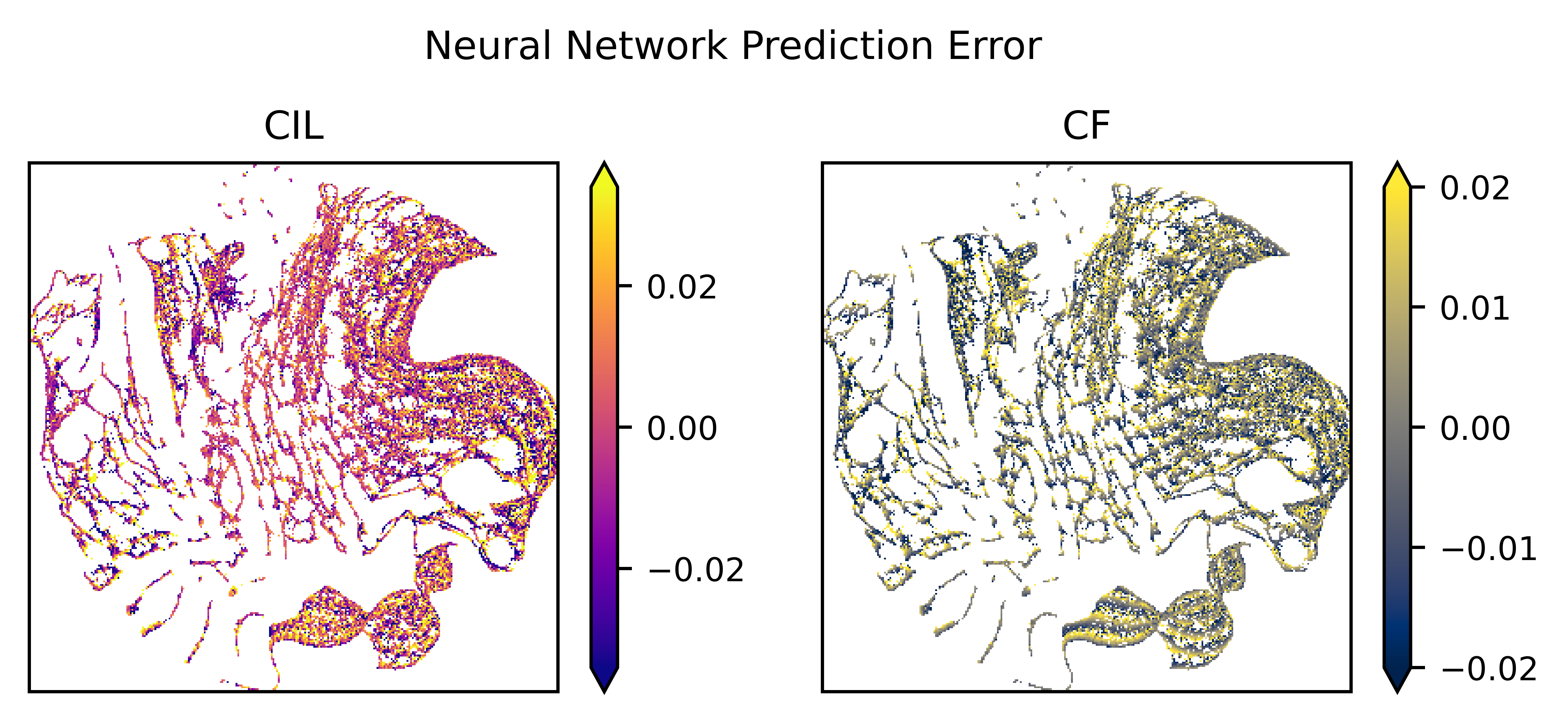}%
\caption{Strand-like regions do not show significantly better prediction accuracy than plane-like regions, even though movies are expected to separate more clearly there. Instead, bands of alternating high and low accuracy are mostly formed between lines of constant $\alpha_{\text{CIL}}$ (spaced 0.1 $\alpha_{\text{CF}}$ apart).
This suggests the network’s loss is generally due to $\alpha_{\text{CF}}$ error.}%
\label{fig:dimensionful_umap_error}%
\end{figure}

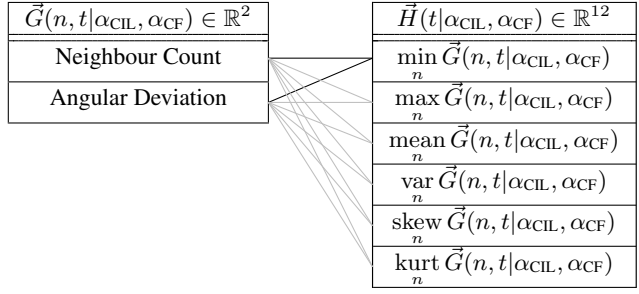
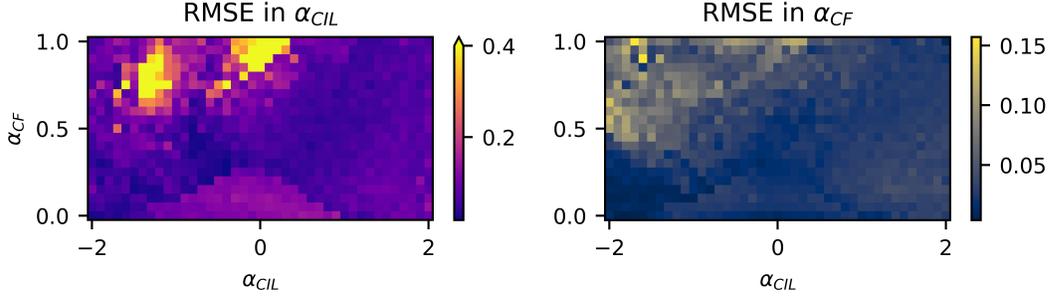
\begin{figure}[H]
\centering
\subfloat[Structural and dynamic cell group features of the $n^{\mathrm{th}}$ cell at time $t$, $\featGfull{}$, coarse-grained over all cells at each time in six different ways to give a 12-dimensional feature vector $\featHfull{}$.]{%
\begin{minipage}{\textwidth}
\footnotesize
\centering
\begin{NiceTabular}[t]{cp{3em}|>$c<$|}
\hhline{-~-}
\multicolumn{1}{|c|}{$\featGfull{} \in \mathbb{R}^{2}$} && \featHfull{} \in \mathbb{R}^{12}\\\hhline{|=|~=}
\multicolumn{1}{|c|}{Neighbour Count}   && \underset{n}{\operatorname{min}}{\, \featGfull{}}\\\hhline{|-|~-}
\multicolumn{1}{|c|}{Angular Deviation} && \underset{n}{\operatorname{max}}{\, \featGfull{}}\\\hhline{|-|~-}
                  && \underset{n}{\operatorname{mean}}{\, \featGfull{}}\\\hhline{~~-}
                  && \underset{n}{\operatorname{var}}{\, \featGfull{}}\\\hhline{~~-}
                  && \underset{n}{\operatorname{skew}}{\, \featGfull{}}\\\hhline{~~-}
                  && \underset{n}{\operatorname{kurt}}{\, \featGfull{}}\\\hhline{~~-}
\CodeAfter
\begin{tikzpicture}
    \draw (2.5-|2) -- (2.5-|3);
    \draw (3.5-|2) -- (2.5-|3);
    \foreach \y in {3.5,...,7.5}{
            \draw[color=black!25] (2.5-|2) -- (\y-|3);
            \draw[color=black!25] (3.5-|2) -- (\y-|3);
    }
\end{tikzpicture}%
\end{NiceTabular}%
\end{minipage}%
\label{tbl:dimless_parameter_prediction}%
}\\
\subfloat[Root mean square error in predicted $\alpha_{\text{CIL}}$ and $\alpha_{\text{CF}}$ from neural network trained on $\featHfull{}$.]{%
\begin{minipage}{\textwidth}
\includegraphics[]{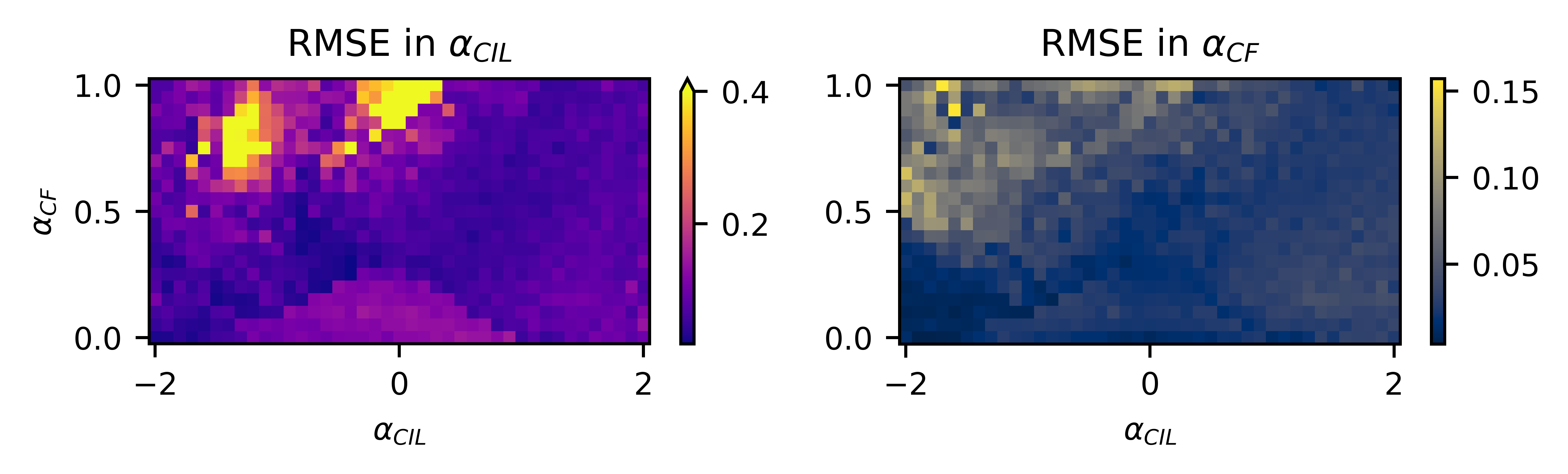}%
\end{minipage}%
\label{fig:dimless_parameter_prediction_rmse}%
}%
\caption{Predictions of $\alpha_{\text{CIL}}$ and $\alpha_{\text{CF}}$ is larger when neural networks are only trained dimensionless features $\featHfull{}$, compared against networks trained on dimensionful features as well in \cref{fig:parameter_prediction}.
Here, an identical neural network architecture is used, but the average root mean squared error in predictions has increased to $0.147$ for $\alpha_{\text{CIL}}$ and $0.0434$ for $\alpha_{\text{CF}}$.}%
\label{fig:dimless_parameter_prediction}%
\end{figure}

\begin{figure}[H]
\centering
\includegraphics[]{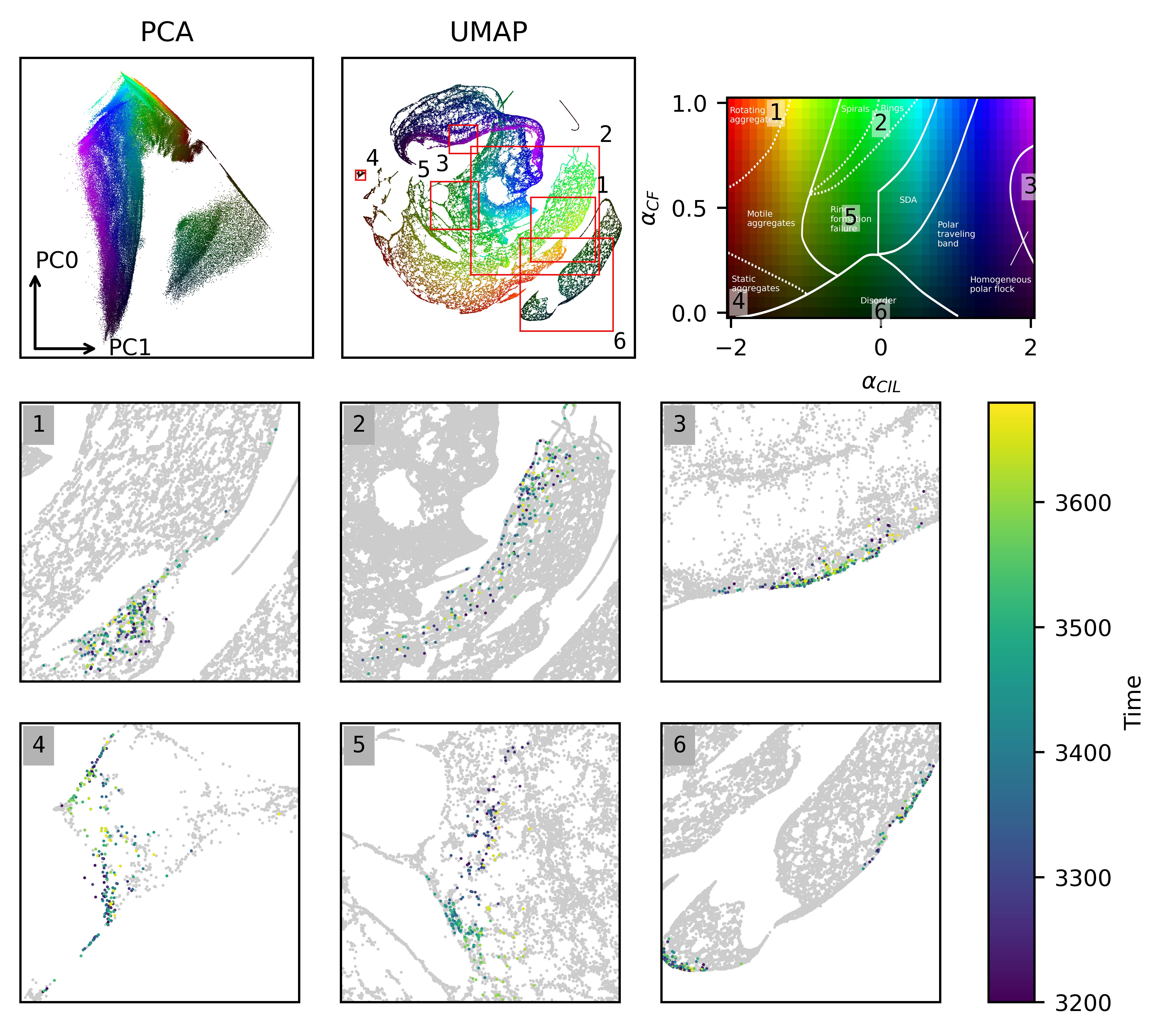}%
\caption{Embedding space learned the neural network in \cref{fig:dimless_parameter_prediction_rmse} using only the dimensionless input-features is more connected compared to that in \cref{fig:dimensionful_embedding_phases}.
Feature vectors of each movie appears also appear more scattered and overlaps more with other movies that have different $\alpha_{\text{CIL}}$ or $\alpha_{\text{CF}}$ parameters, highlighting the importance of the dimensionful features in identifying the parameters.}%
\label{fig:dimensionless_embedding_phases}%
\end{figure}

\begin{figure}[H]
\includegraphics[]{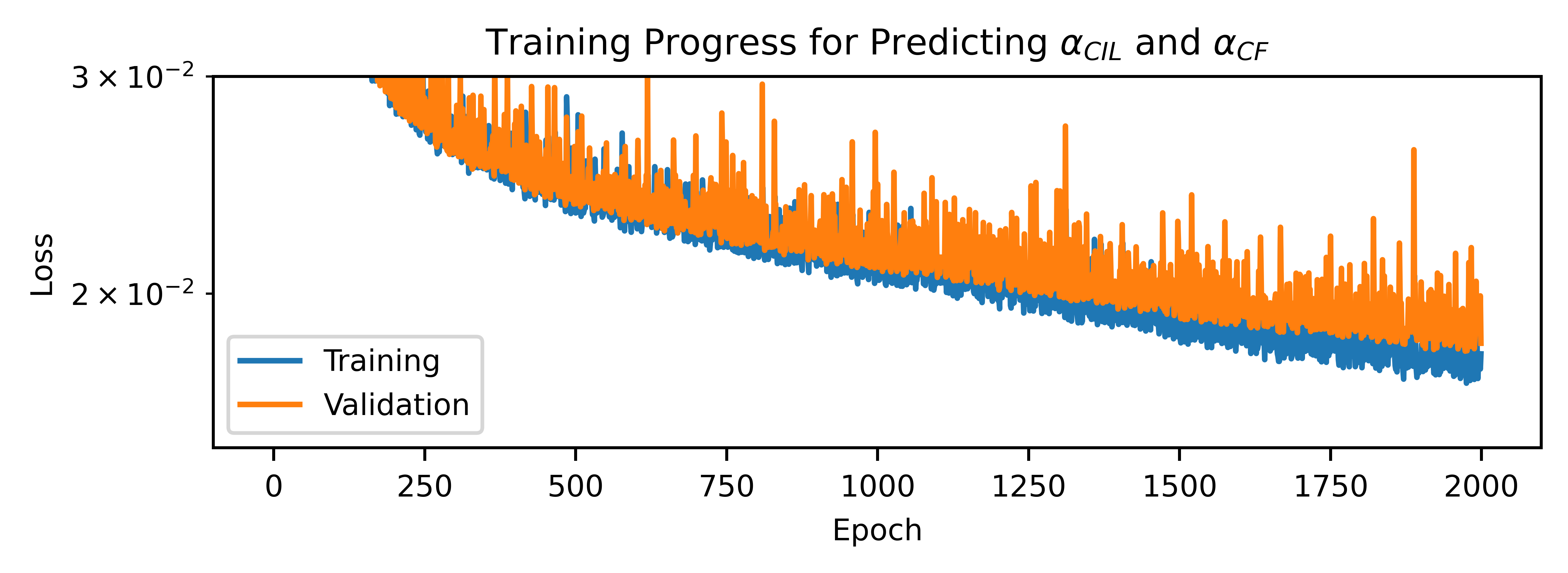}%
\caption{Neural network trained only on dimensionless features (neighbour count and angular deviation) shows comparatively higher mean squared error than the network in \cref{fig:dimensionful_param_loss}, despite the networks having identical architectures.
This highlights the importance of the dimensionless features in predicting the $\alpha_{\text{CIL}}$ and $\alpha_{\text{CF}}$ parameters.}%
\label{fig:dimensionless_param_loss}%
\end{figure}

\begin{figure}[H]
\centering
\includegraphics[]{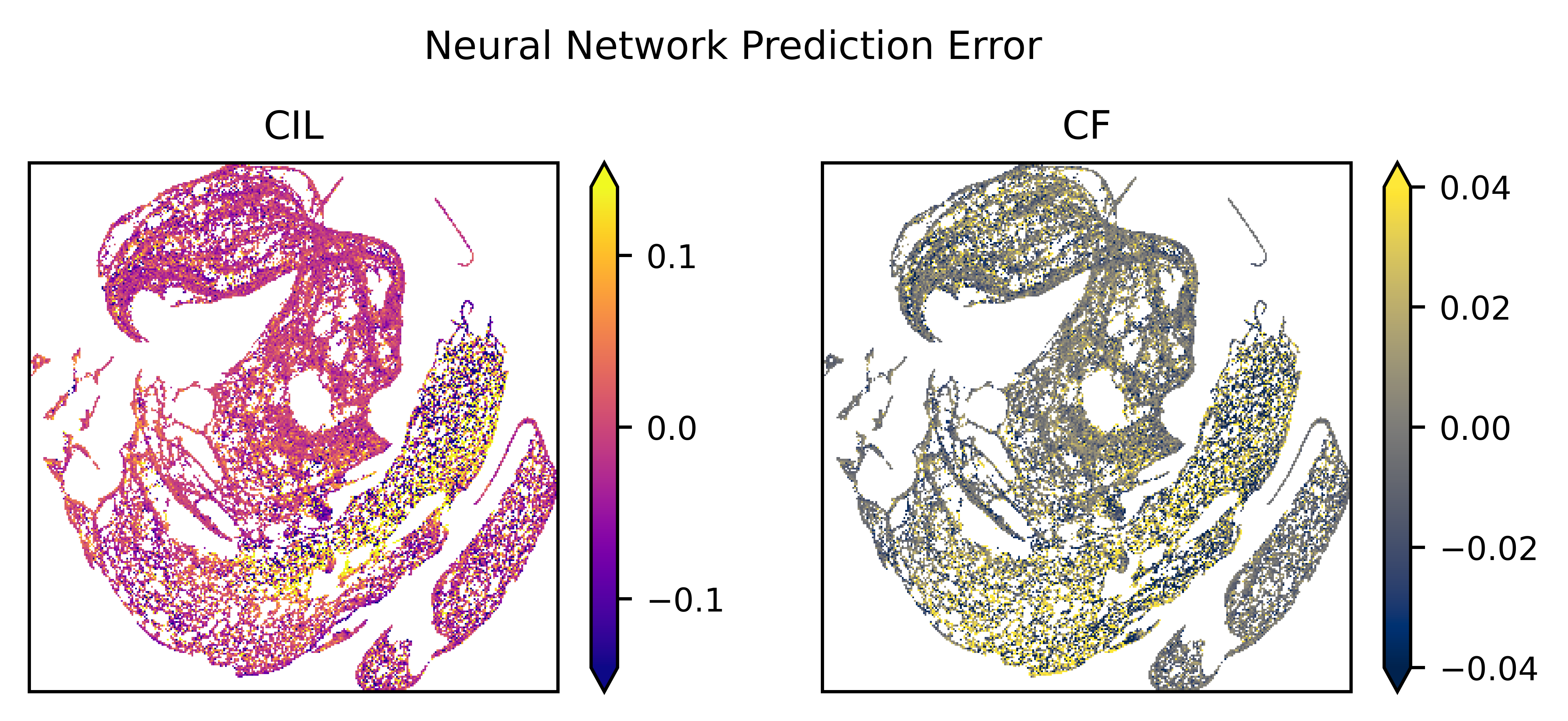}%
\caption{UMAP projected embedding space of the neural network in \Cref{fig:dimensionless_embedding_phases} trained on dimensionless motif features. The $\alpha_{\text{CIL}}$ and $\alpha_{\text{CF}}$ parameter prediction error here is often enough for features from different simulation parameters to be confused and overlap one another. This increase in overlap likely stopped the bands seen in \cref{fig:dimensionful_umap_error} from appearing.}%
\label{fig:dimensionless_umap_error}%
\end{figure}